\documentclass[aps,
prl,
reprint,
twocolumn,
superscriptaddress
]{revtex4-2}

\usepackage{amsmath,amssymb}
\usepackage{graphicx}
\usepackage{float}
\usepackage{natbib}
\usepackage{color}
\usepackage{xcolor}
\usepackage{bbold} 
\usepackage{microtype} 
\usepackage{lmodern}
\usepackage[colorlinks=true,bookmarks=false,citecolor=blue]{hyperref}

\graphicspath{ {./Figures/} }

\newcommand{\eps}{\varepsilon }
\newcommand{\br}{\mathbf{r} }

\newcommand{\bE}{\mathbf{E} }

\newcommand{\bB}{\mathbf{B} }

\newcommand{\un}{\widehat{\mathbf{n}} }

\newcommand{\bbE}{ \mathbb{E} }

\renewcommand{\Im}{\frak{I}\mathrm{m} }

\newcommand{\dyad}[1] {\overset\leftrightarrow{#1}}

\newcommand{\wcsq}{\left(\frac \omega{c}\right)^2 }

\begin{document}

\title{Chiral photonic super-crystals based on helical van der Waals homostructures}

\newcommand{\dipc}{
Donostia International Physics Center (DIPC), Donostia, San-Sebastian, Spain
}

\newcommand{\moscow}{
Moscow Center for Advanced Studies, Moscow, 123592, Russia
}

\newcommand{\ibhf}{
Emanuel Institute of Biochemical Physics RAS, Moscow 119334, Russia
}

\newcommand{\xpan}{
Emerging Technologies Research Center, XPANCEO, Dubai Investment Park First, Dubai, UAE
}

\newcommand{\yerevan}{
Laboratory of Advanced Functional Materials, Yerevan State University, Yerevan, 0025, Armenia
}

\newcommand{\novoselov}{
National Graphene Institute (NGI), University of Manchester, Manchester, M13 9PL, UK
}

\newcommand{\novoselovv}{Department of Materials Science and Engineering, National University of Singapore, Singapore, 03-09 EA, Singapore}

\newcommand{\novoselovvv}{Chongqing 2D Materials Institute, 400714, Chongqing, China}

\author{Kirill V. Voronin}
\thanks{These authors contributed equally to this work}
\affiliation{\dipc}

\author{Adilet N. Toksumakov}
\thanks{These authors contributed equally to this work}
\affiliation{\ibhf}
\affiliation{\moscow}

\author{Georgy A. Ermolaev}
\thanks{These authors contributed equally to this work}
\affiliation{\xpan}

\author{Aleksandr S. Slavich}
\affiliation{\moscow}

\author{Mikhail K. Tatmyshevskiy}
\affiliation{\moscow}

\author{Sergey V. Novikov}
\affiliation{\moscow}

\author{Andrey A. Vyshnevy}
\affiliation{\moscow}
\affiliation{\xpan}

\author{Aleksey V. Arsenin}
\affiliation{\xpan}
\affiliation{\yerevan}

\author{Kostya S. Novoselov}
\affiliation{\novoselov}
\affiliation{\novoselovv}
\affiliation{\novoselovvv}

\author{Davit A. Ghazaryan}
\email[]{dav280892@gmail.com}
\affiliation{\moscow}
\affiliation{\yerevan}

\author{Valentyn S. Volkov}
\affiliation{\xpan}
\affiliation{\yerevan}

\author{Denis G. Baranov}
\email[]{baranov.mipt@gmail.com}
\affiliation{\moscow}

\begin{abstract}
Chirality is probably the most mysterious among all symmetry transformations. Very readily broken in biological systems, it is practically absent in naturally occurring inorganic materials and is very challenging to create artificially. Chiral optical wavefronts are often used for the identification, control and discrimination of left- and right-handed biological and other molecules. Thus, it is crucially important to create materials capable of chiral interaction with light, which would allow one to assign arbitrary chiral properties to a light field. In this paper, we utilized van der Waals technology to assemble helical homostructures with chiral properties (e. g. circular dichroism). Because of the large range of van der Waals materials available such helical homostructures can be assigned with very flexible optical properties. We demonstrate our approach by creating helical homostructures based on multilayer As$_2$S$_3$, which offers the most pronounced chiral properties even in thin structures due to its strong biaxial optically anisotropy. Our work showcases that the chirality of an electromagnetic system may emerge at an intermediate level between the molecular and the mesoscopic one due to the tailored arrangement of non-chiral layers of van der Waals crystals and without additional patterning.
\end{abstract}

\maketitle

\section{Introduction}

A geometrical object in three-dimensional space is called chiral if it cannot be aligned with its mirror image by a series of rotations and translations \cite{Kelvin1894}. Chirality occurs at various scales ranging from the shapes of galaxies down to numerous bio-molecules \cite{chhabra2013review} and quantum electronic states \cite{asenjo2014dichroism, Wallbank2016, Stickler2021}. 
In nanophotonics, the optical chirality at the qualitative level affects the interactions of matter with electromagnetic radiation \cite{Lindell2018, Inoue2004}. In particular, the transmission of a chiral (circularly-polarized) electromagnetic field through a chiral sample results in circular dichroism, which underlies numerous techniques of discriminating molecular enantiomers \cite{Govorov2010, Tang2010, Hendry2010}. Geometrically chiral arrangements of dielectric and plasmonic meta-atoms into metasurfaces have enabled highly flexible manipulation of optical radiation \cite{Gorkunov2020,chen2023observation}, chiral mirrors \cite{plum2015chiral}, and handedness-preserving optical cavities \cite{Feis2020, Voronin2022, schafer2023chiral, PhysRevA.107.L021501}. 

The design of various chiral optical nanostructures and metasurfaces \cite{han2023recent} typically involves achiral media. They possess the so-called \emph{configurational} chirality, wherein the chirality of an object originates from the geometric arrangements of its components, rather than the microscopic chirality of the underlying media \cite{hentschel2012three, schaferling2012tailoring,  avalos2022chiral, Cecconello2017}. Recent studies successfully demonstrated chiral structures based on twisted stacking of otherwise non-chiral materials, such as graphene \cite{kim2016chiral} and h-BN \cite{ochoa2020flat}. The chirality results from the helical symmetry of moiré structures, where the crystal lattices of adjacent atomic layers are twisted and crystallographically aligned with respect to each other up to several degrees. Another good demonstration of such twisted systems composed of non-chiral materials is the homo- and hetero-bilayers of transition metal dichalcogenides, such as MoSe$_2$-WSe$_2$ or WSe$_2$-WSe$_2$ \cite{Wu2022, forg2021moire, Chen2022}. Here, their crystal lattices are either twisted and crystallographically aligned with respect to each other by several or even tens of degrees, in both cases leading to the exceptionally strong normalized chiroptical response of their entire moiré structures, which however, is of different physical origin and arises from microscopic mechanism based on transition metal dichalcogenides valley selection rule.

Leveraging the strong normalized chiroptical response of twisted stacks to achieve high absolute circular dichroism is challenging since it would require multiple repeated steps of stacking 2D crystals on top of each other. Here, one might opt for stacking a few optically thick layers of their bulk counterparts - van der Waals (vdW) crystals. However, since the large part of vdW crystals are in-plane isotropic crystals \cite{li2014measurement, mak2016photonics}, it would render the entire structure non-chiral due to the loss of helical symmetry. In this regard, recently discovered \emph{biaxial} vdW crystals with a strong in-plane anisotropy may offer new possibilities \cite{khaliji2022twisted, zhang2022photonic}. These materials feature \emph{biaxial} permittivity tensors due to the low symmetry of their atomic lattices \cite{Siskins2019, Shubnic2020, alvarez2020infrared}. This sophisticated anisotropy enables complex optical phenomena, such as hyperbolic dispersion \cite{ma2018}, anomalous refraction \cite{Duan2021}, canalization \cite{Hu2020, Duan2020}, and many others. Despite the diversity of possible anisotropic optical properties, bulk vdW crystals are typically achiral. However, as we will show below, a few cleverly arranged in-plane anisotropic vdW layers form a geometrically chiral system with a strong chiroptical response that can be easily fabricated.

In this work, we devise a theory for helical photonic super-crystals based on the stack of vdW crystal layers, where each of the constituent layer possesses a \emph{biaxial} anisotropy. We find the bandstructure of a periodic system, identify handedness-dependent bandgaps in its spectrum, and observe circular dichroism in such truncated chiral photonic super-crystals. Furthermore, based on our theoretical predictions, we demonstrate the emergence of optical chirality in helical vdW homostructures of As$_2$S$_3$ layers arranged in a chiral fashion. Owing to a strong in-plane anisotropy, As$_2$S$_3$ serves as a perfect material platform for the observation of the predicted effects when assembled into a helical trilayer homostructure \cite{slavich2023exploring}. In particular, it allows the observation of the circular dichroism employing a unique technique based on imaging spectroscopic ellipsometry. Our work showcases that the geometrical chirality of an electromagnetic system may emerge at an intermediate level between the molecular and the mesoscopic one due to the tailored arrangement of non-chiral layers of vdW crystals without the introduction of additional patterning.

\section{Results}

\begin{figure}[t!]
\centering\includegraphics[width=0.5\textwidth]{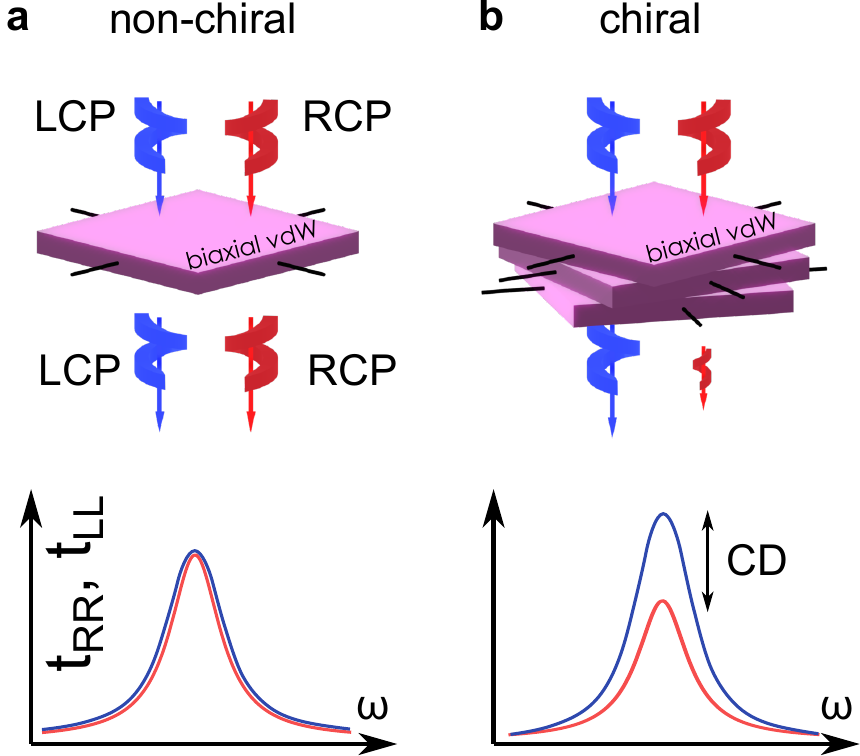}
\caption{\textbf{Emergence of chirality in a helical vdW super-crystal.} (a) A single optically thick layer of vdW crystal may possess a biaxial  anisotropy, yet it remains achiral demonstrating zero circular dichroism. (b) A homostructure made of a few identical vdW layers specifically oriented in a certain fashion possesses a configurational chirality, which gives rise to circular dichroism.}
\label{fig0}
\end{figure}

A single optically thick layer of a biaxial vdW crystal is optically anisotropic and exhibits the associated phenomena of linear dichroism, but nonetheless, it is geometrically achiral and it does not feature any circular dichroism, Figure \ref{fig0}(a).
However, when a few identical vdW layers are combined in a helical homostructrue, the entire crystal acquires geometrical chirality giving rise to an observable circular dichroism, Figure \ref{fig0}(b).

\subsection{Infinite periodic super-crystal}

\begin{figure*}[hbt!]
\centering\includegraphics[width=0.8\textwidth]{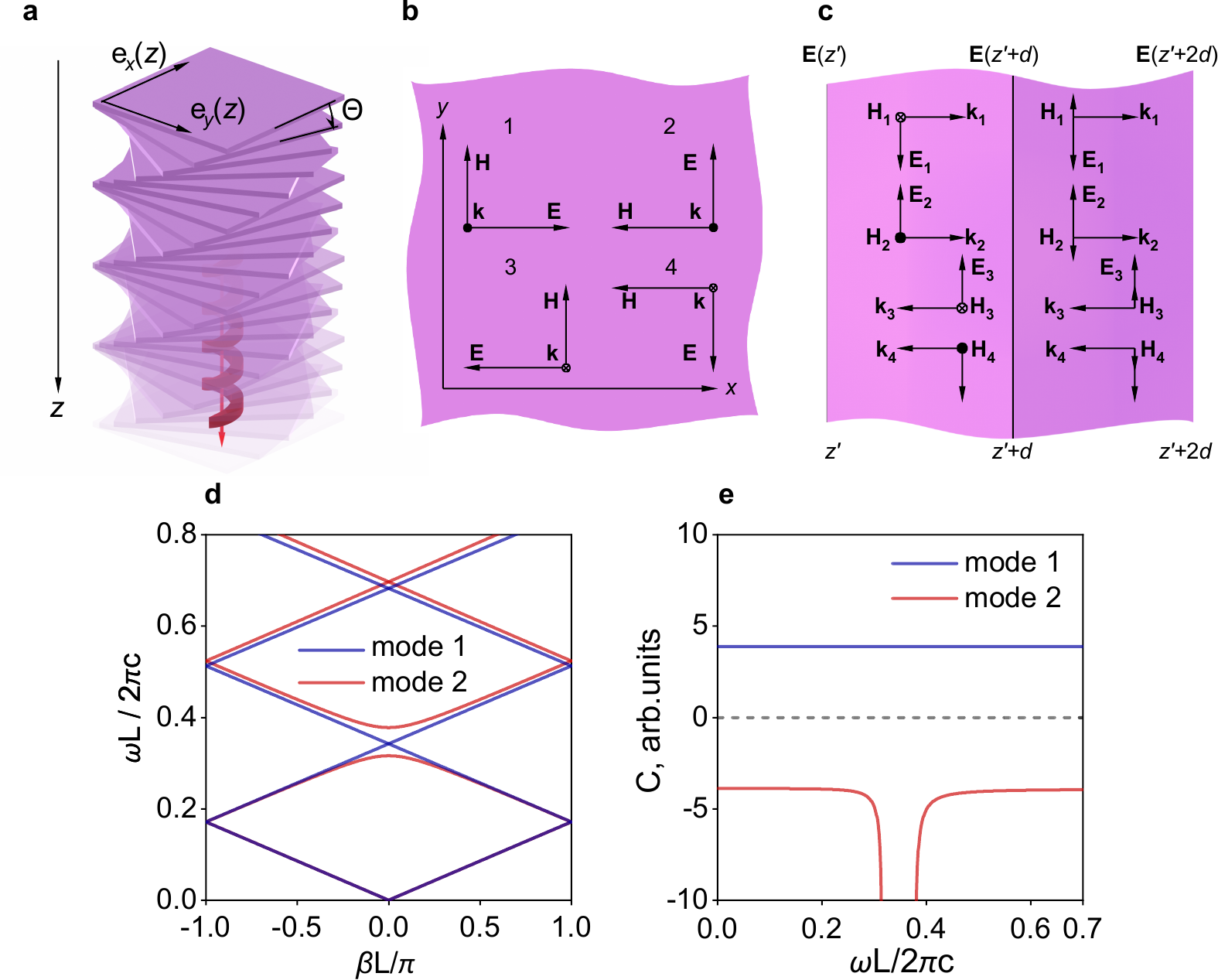}
\caption{\textbf{Infinite photonic chiral super-crystal.} (a) Sketch of the system under the study. (b) Basis of the modes in the main axes of each layer. (c) Field decomposition on the basis of neighboring layers. (d) The spectrum of the Bloch modes of a chiral super-crystal with $\eps_x=10$, $\eps_y=7$.
(e) Optical chirality density $C$ of two branches of the infinite periodic chiral super-crystal as in panel (d) versus normalized frequency. The discontinuity of mode 1 matches the bandgap in the spectrum of the extended modes.}
\label{fig1}
\end{figure*}

We begin by examining a periodic infinite stack of anisotropic biaxial layers shown in Figure \ref{fig1}(a). The unit cell consists of $N$ identical parallel layers of thickness $d$ but oriented in different ways.
In an appropriate Cartesian basis, the permittivity tensor of the material reads 
\begin{equation}
    \dyad{\eps} = \mathrm{diag} (\eps_x, \eps_y, \eps_z),
\end{equation}
where $\eps_{x} \ne \eps_{y} \ne \eps_{z} \ne \eps_{x}$ rendering such a material biaxial \cite{new2013biaxial}. Without loss of generality, we assume $\eps_x > \eps_y$.
For all layers, the $z$-axis is perpendicular to the surfaces of the layers, and $x$ and $y$ denote the in-plane principal axes of the material in the respective layer.

The layers are arranged in such a way that the $x$ optical axis of $n$-th layer in the unit cell makes an angle $\theta$ with the axis $x$ of $(n-1)$-th layer in the unit cell as shown in Figure \ref{fig1}(a).
The permittivity tensor of $n$-th layer takes the form
\begin{equation}
    \dyad{\eps}_n = \dyad{R}^{-1}_n \dyad{\eps} \dyad{R}_n,
\end{equation}
where $\dyad{R}_n$ is an in-plane rotation matrix for $n$-th layer (see Supporting Section S1 for more details). Unless $\theta = \pm \pi, \pm \pi/2$, this arrangement of anisotropic layers breaks all possible mirror and inversion symmetries of the system and renders the entire stack geometrically chiral in the absence of intrinsic chirality of each of the constituent layers.

The structure forms a photonic crystal with the lattice constant $L = N d$ and the Brillouin zone $\left[- G/2, G/2 \right]$ with $G = 2\pi/L = \theta/d$ being the reciprocal lattice constant \cite{Molding1995photonic}. In the following, we limit our analysis to the case of on-axis propagation, $k_x = k_y = 0$. The case of off-axis propagation can be analyzed analogously. In each layer, we expand the field by the basis consisting of the waves propagating along or against $z$-axis and polarized along $x$- and $y$-axes, as shown in Figure \ref{fig1}(b):
\begin{align}
    \bE(\br) = E_1 (z) \un ' + E_2 (z) \un '' - E_3 (z) \un' - E_4 (z) \un '',
    \label{Eq_E_expansion}
\end{align}
where $E_i(z)$ are yet unknown complex field amplitudes, and $\un'$ and $\un''$ are unit vectors directed along the $x$ and $y$ principal axes of the corresponding layer. The dependence of $E_i$ on $z$ expresses the fact that the amplitudes of the four basic waves evolve across each layer and from layer to layer.

To facilitate the analytical derivation of the band structure of the chiral super-crystal, let us analyze a special case of subwavelength layers, $\sqrt{\varepsilon_j} d \ll \lambda$. Then the photonic crystal can be treated as a continuous (but non-uniform) medium.
Imposing boundary conditions on the total field, we establish a linear system of the differential equations governing the spatial evolution of the complex amplitudes of waves $E_1$, $E_2$, $E_3$, and $E_4$ (see Supporting Section S1):
\begin{equation}
    \frac{d}{d z} \mathbb{E} = \dyad{ \mathbb{U}} \mathbb{E},
    \label{Syst}
\end{equation}
where $\mathbb{E} = (E_1, E_2, E_3, E_4)^T$ is the four-dimensional vector of complex amplitudes, and $\dyad{ \mathbb{U}}$ is a linear operator dictating the spatial evolution of the four waves (see Supporting Section S1). The solution of the system (\ref{Syst}) is given by:
\begin{equation}
   \mathbb{E}(z) = \bbE_1^+ e^{i \beta_{1} z} + \bbE_2^+ e^{i \beta_{2} z} + \bbE_1^- e^{-i \beta_{1} z} + \bbE_2^- e^{-i \beta_{2} z},
    \label{Sol}
\end{equation}
where $\bbE_{1,2}^{\pm}$ are the eigenvectors of the system (\ref{Syst}) and $\pm i \beta_{1,2}$ are the eigenvalues of the system (\ref{Syst}), which are given by:
\begin{equation}
    \beta_{1,2}=\sqrt{G^2 + \overline{\eps} \wcsq \pm \frac{\omega}{c} \sqrt{4 \overline{\eps} G^2 + \frac{\Delta \eps^2 }{4} \wcsq }},
    \label{PCWV}
\end{equation}
where $\overline{\eps} = (\varepsilon_x + \varepsilon_y)/2$ is the average in-plane permittivity, and $\Delta \varepsilon = \varepsilon_x - \varepsilon_y$ is the in-plane permittivity asymmetry.

\begin{figure*}[t!]
\centering
\includegraphics[width=0.9\textwidth]{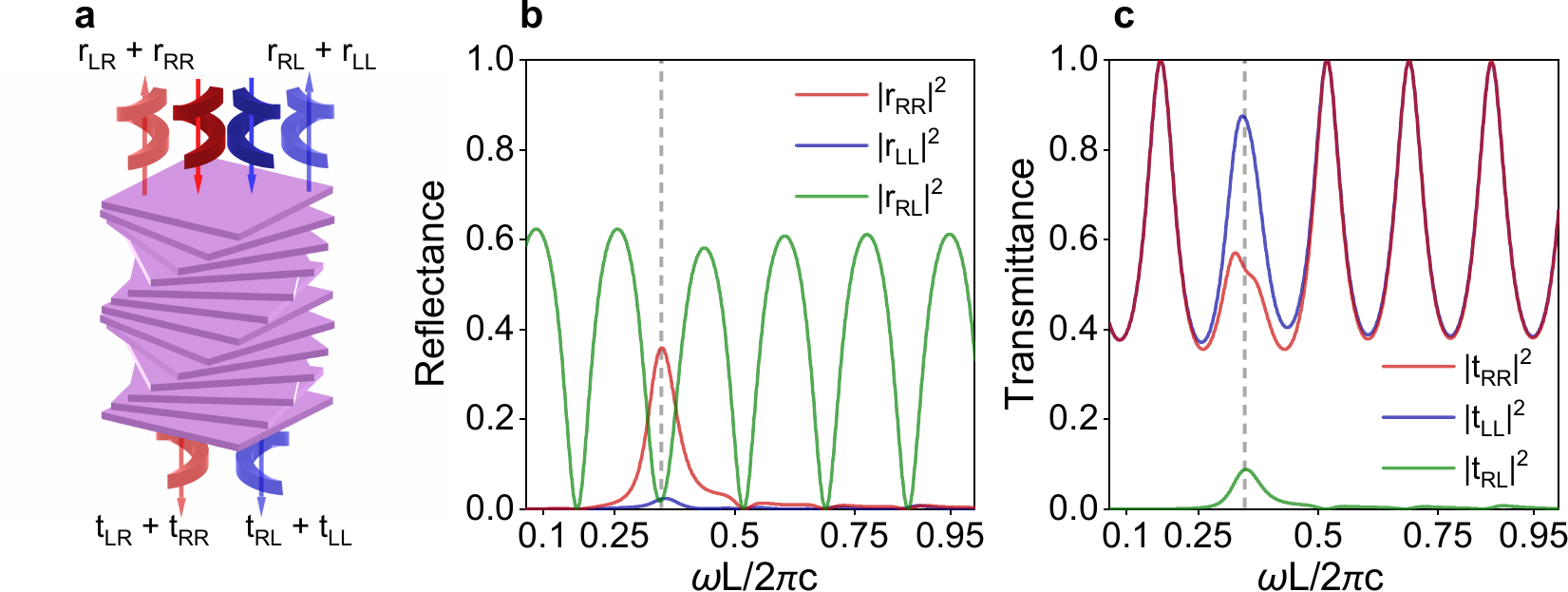}
\caption{\textbf{Scattering properties of a finite chiral super-crystal}. (a) Sketch of the reflection and transmission of the finite chiral system. (b) Reflection and (c) transmission amplitudes for a finite structure consisting of one unit cell of the photonic crystal, that is, one complete rotation of the main axis, as a function of its thickness $d$ for $\varepsilon_x = 10$, $\varepsilon_y = 7$, $\varepsilon_1 = \varepsilon_2 = 1$.}
\label{fig2}
\end{figure*}

{The solution expressed by Eq. \eqref{Sol} is a linear combination of two forward- and two backward-propagating linearly-independent Bloch modes with quasi-momenta defined up to an integer of the reciprocal lattice constant $\pm \beta_{1,2} + G m$, $m \in \mathbb{Z}$.}
Note that Eq. \eqref{Sol} expresses a four-dimensional vector of amplitudes $(E_1,E_2,E_3,E_4)^T$. The coordinate dependencies of the Cartesian components of electric and magnetic fields of each Bloch mode are found by substituting the eigenvectors $\bbE_{1,2}^{\pm}$ into Eq. \eqref{Eq_E_expansion}.
We emphasize that this solution was obtained in the approximation of optically thin layers. The unit cell itself does not have to be optically thin for this approximation to hold: it is only the quasi-continuous variation of the orientation angle $\theta$ that matters for our derivation.

Figure \ref{fig1}(d) shows an exemplary spectrum of Bloch modes of a chiral super-crystal with $\overline{\eps} = 8.5$, $\Delta \varepsilon = 3$. These values, taken as an example, are close to the parameters of natural biaxial material As$_2$S$_3$ in the visible range \cite{slavich2023exploring}, which allows the experimental realization of the introduced systems. 
The structure used for calculations in the following possesses a right-handed (RH) geometric chirality: the stack features a right-handed helix in space with a clockwise forward rotation when viewed along the axis (either from top or bottom of the structure).
Both modes start from the $\Gamma$-point of the periodic system, $\beta = 0$, and follow nearly the same dispersion. 
However, as the frequency approaches a certain value the spectrum of the chiral super-crystal exhibits the most remarkable feature: while the dispersion of the mode '1' remains continuous, that of the mode '2' experiences a bandgap.
By equating the expression for the Bloch momentum \eqref{PCWV} to zero, we obtain the lower and the upper edges of the band gap: $1/\sqrt{\varepsilon_x} \le \omega L/ (2 \pi c) \le 1/\sqrt{\varepsilon_y}$ (see Supporting Section S1).

Electromagnetic handedness of propagating modes is quantified by calculating their local chirality density \cite{Lipkin1964, Tang2010}:
\begin{equation}
    C(\br,\omega) = \frac{\eps_0\omega }{2} \Im[\bE\cdot \bB^*].
    \label{Eq_8}
\end{equation}
We refer to the Bloch modes with $C < 0$ as the right-handed (RH) ones and the ones with $C > 0$ as the left-handed (LH) ones.
Figure \ref{fig1}(e) shows the resulting optical chirality density $C$ of both Bloch modes versus normalized frequency. 
While mode '1' possesses positive chirality density associated with the LH field, mode '2' possesses negative chirality density associated with the RH field, and features a gap around $\omega L/ (2 \pi c) \approx 0.35$.
{Clearly, inverting the geometric handedness of the super-crystal flips this behavior of the Bloch modes.}


Now we assume that the entire unit cell of the super-crystal is subwavelength, $N d \ll \lambda$. The photonic crystal then can be treated as a uniform medium according to the effective
medium approximation, which can be described by the constitutive relations of \textit{bi}-isotropic chiral medium, characterized by the effective dielectric permittivity and the Pasteur parameter, which are given by the following expressions (see Supporting Section S2 for more details):
\begin{equation}
    \sqrt{\eps_{eff}} = \sqrt{\overline{\eps}} + \frac{\Delta \varepsilon^2}{32 G^2 \sqrt{\overline{\eps}}} \wcsq, \
    \kappa_{eff} = \frac{\Delta \eps^2 }{32 G^3} \left( \frac{\omega}{c} \right)^3.
    \label{Epseff}
\end{equation}

Expectedly, the in-plane permittivity asymmetry favors the effective Pasteur parameter $\kappa_{eff}$ and vanishes for isotropic films with $\Delta \eps = 0$.
What is far less intuitive, however, is that the same effective Pasteur parameter decreases with the increase of the screw rate of the multilayer stack $G \propto 1/L$. One could expect that the higher rotation rate of the helical structure leads to a larger degree of chirality. However, according to Eq. \eqref{Epseff}, the Pasteur parameter is in inverse proportionality with the rotation rate. Qualitatively it can be understood by appreciating that the longer the wavelength of light compared to the length of the coil, i.e. the cell size of a photonic crystal, the more uniform the field is on the scale of one turn, the less influence the twist has on the wave propagation. This is in agreement with the informal statement that "chirality needs volume" \cite{caloz2020electromagnetic}.

\begin{figure*}[t!]
\centering\includegraphics[width=.8\textwidth]{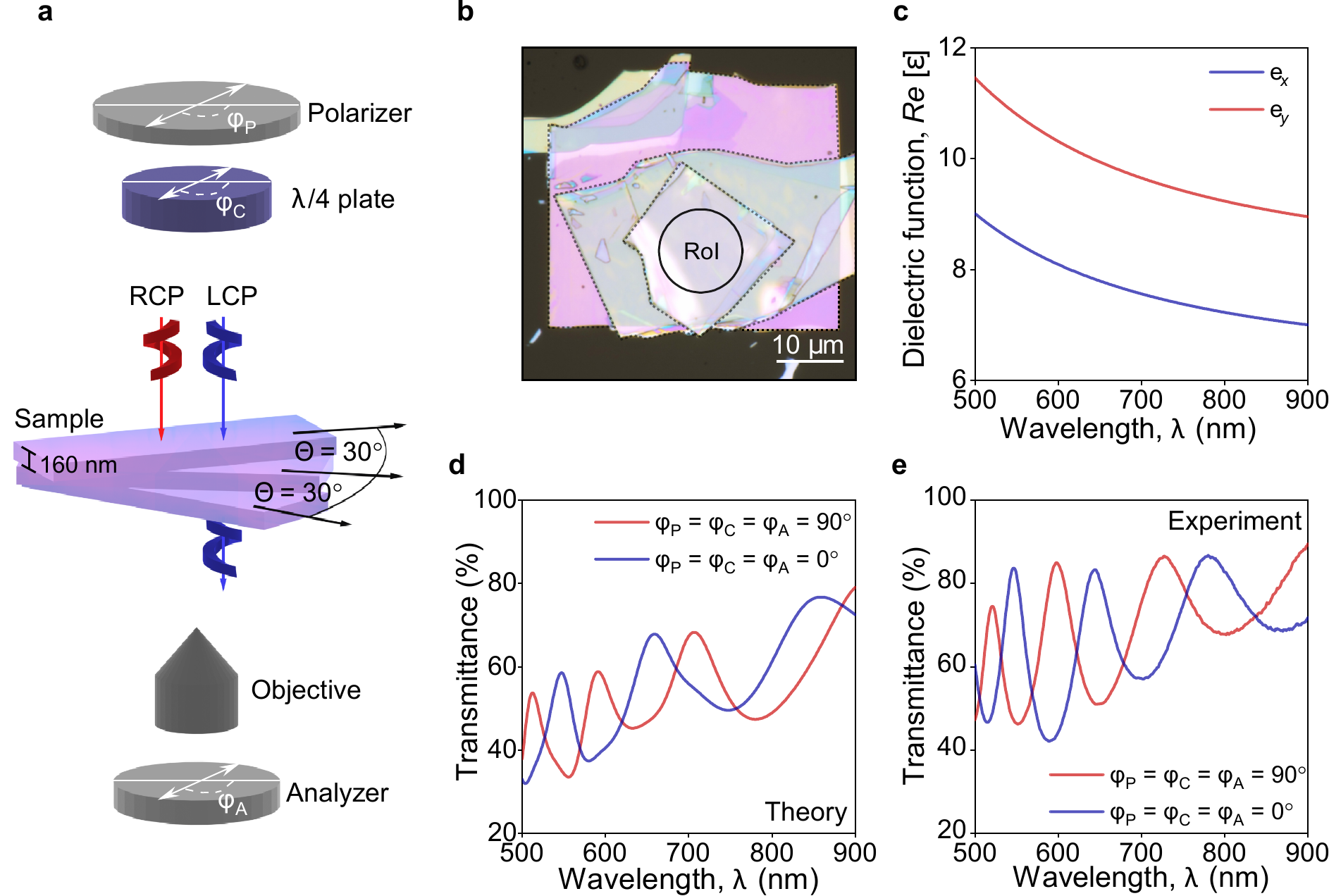}
\caption{\textbf{Chiral photonic super-crystal based on helical As$_2$S$_3$ van der Waals homostructure, initial characterization.} (a) Schematic representation of the experimental setup. RCP stands for right circular polarized light, and LCP for left circular polarized light. (b)  Optical micrograph of the fabricated chiral super-crystal based on helical vdW homostructure of trilayer As$_2$S$_3$. The edges of the layers are highlighted by dashed lines. (c) Dielectric function of a separate bulk vdW As$_2$S$_3$ layer. (d) Theoretically calculated transmittance spectra for horizontal (H or 0°) and vertical (V or 90°) orientations of the polarizer, compensator, and analyzer. (e) Same as (d), but obtained experimentally.}
\label{fig3}
\end{figure*}

\subsection{Finite chiral super-crystal}

Next, we examine the optical scattering response of a truncated chiral super-crystal. To that end, we consider the structure placed between two isotropic half-spaces with dielectric permittivities $\varepsilon_1$ and $\varepsilon_2$. 
Reflection and transmission coefficients of the finite super-crystal can be obtained by applying the same formalism of transfer matrices, but this time without the need to impose the Bloch wave periodic conditions. The details of the calculations of the finite system are summarized in Supporting Section S3. In this section, we consider the structure placed between two identical media, $\varepsilon_1 = \varepsilon_2 = 1$.


{The relationship between the amplitudes of the incident and scattered waves in the base of circularly polarized plane waves is described by the scattering matrix $\dyad{\mathbb{S}}$ (see Supporting Section S3). 
In the following, $r_{\mu \nu}$ ($r_{\mu \nu}'$) denote complex reflection coefficients on the side 1(2) from the polarization state $\nu$ into polarization state $\mu$, 
and $t_{\mu \nu}$ ($t_{\mu \nu}'$) denote complex transmission coefficients for light incident from side 1(2) in polarization state $\nu$ to the polarization state $\mu$ on the side 2(1).}
The presence of non-zero elements in the $S$-matrix depends on the symmetries of the system.
For a highly symmetric structure made of isotropic films all the elements of this matrix are zero except for co-polarized transmission $t_{\mu \mu}$ and $t'_{\mu \mu}$, and cross-polarized reflection $r_{\mu \nu}$ and $r'_{\mu \nu}$ ($\mu \ne \nu$) amplitudes.
Violating different symmetries of the structure allows certain matrix elements to be non-zero \cite{Menzel2010}. Since our finite super-crystal does not, generally, possess any out-of-plane mirror symmetries or vertical axes of rotation of order $n \ge 3$, all elements of the scattering matrix, are, generally, non-zero.  

\begin{figure*}[hbt!]
\centering\includegraphics[width=.8\textwidth]{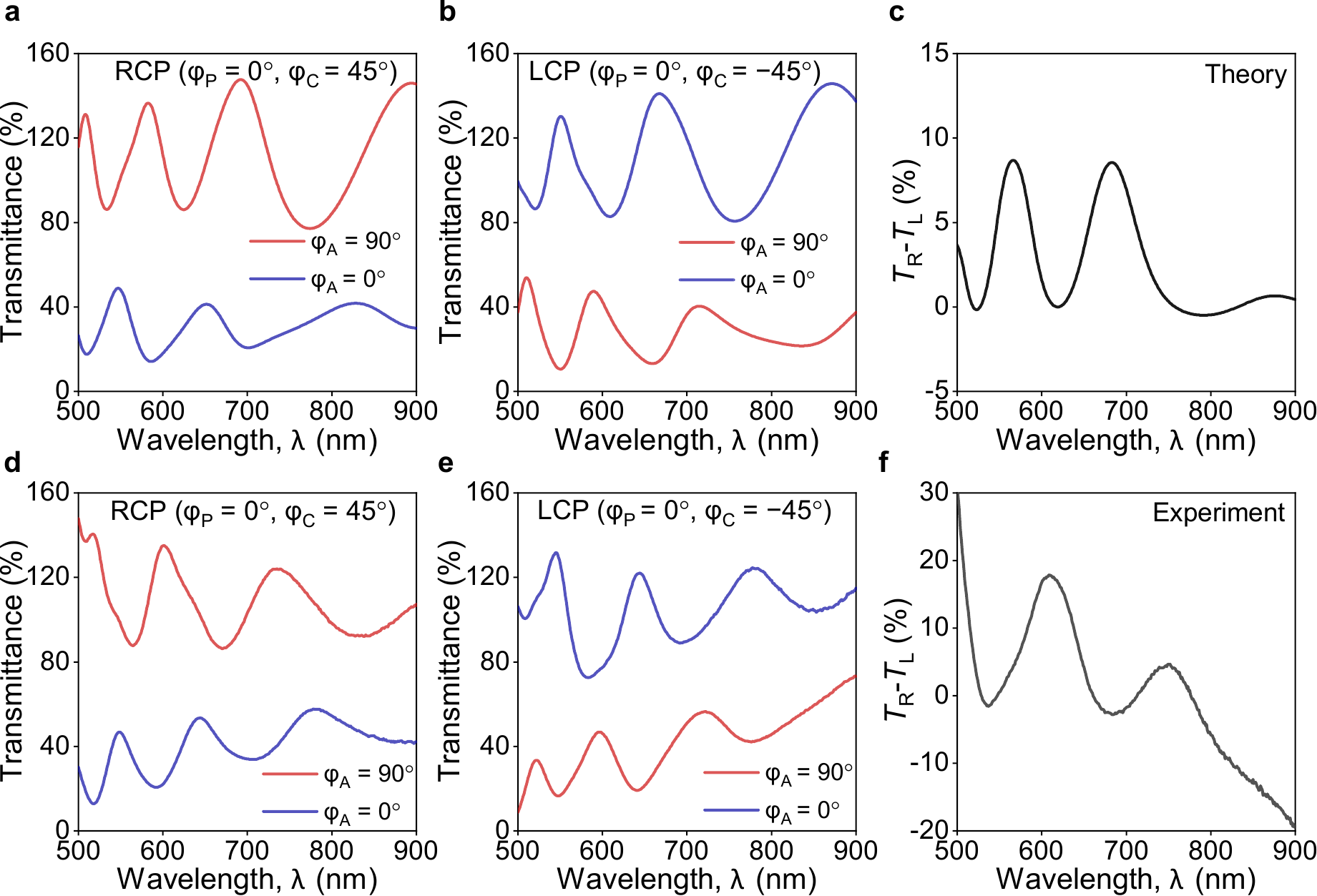}
\caption{\textbf{Emerging optical chirality in helical van der Waals homostructures.} (a) Calculated transmission of RCP light at horizontal ($\varphi_A$ = 0°) and vertical ($\varphi_A$ = 90°) orientations of the analyzer. (b) Same as (a), but for LCP light. (c) Calculated difference between $T_R$ and $T_L$ signals. (d) Experimentally measured transmittance spectra for RCP at horizontal ($\varphi_A$ = 0°), and vertical ($\varphi_A$ = 90°) orientations of the analyzer. (e) Same as (d), but for LCP light. f) Experimentally obtained difference between $T_R$ and $T_L$.}
\label{fig4}
\end{figure*}

This is readily appreciated in Figure \ref{fig2}(b,c), which shows the elements of the scattering matrix of a finite, containing one turn of the helix, chiral super-crystal at normal incidence excitation for $\varepsilon_x = 10$, $\varepsilon_y = 7$, $\varepsilon_1 = \varepsilon_2 = 1$. 
Co-polarized transmission $t_{\mu \mu}$ and cross-polarized reflection $r_{\mu \nu}$ spectra exhibit an expected series of Fabry-Perot resonances of the dielectric stack.
Only around the band gap of the corresponding infinite structure, $\omega L/ (2 \pi c) \approx 0.35$, substantial circular dichroism $|t_{RR}| \ne |t_{LL}|$ comes into play.
The low symmetry of the structure at the same time results in unequal co-polarized reflection around the band gap, $r_{RR} \ne r_{LL}$. Cross-polarized reflections remain equal due to the reciprocity of the system, $r_{RL} = r_{LR}$.
The only symmetry the finite super-crystal maintains is the presence of a horizontal axis of rotation by $\pi$. Application of this transformation to the system permutes the sides but does not affect the handednesses of the incident and scattered waves, which along with reciprocity results in the identity $|t_{RL}| = |t_{LR}|$, which is confirmed by the calculated spectra in Figure \ref{fig2}(c).

Figure S4 of Supporting Information shows more examples of polarization-resolved reflection and transmission spectra for a structure made of vdW films with more prominent (somewhat exaggerated) in-plane anisotropy with $\eps_x = 10$ and $\eps_y = 2$.
A remarkable behavior is observed at $\omega L/ (2 \pi c) \approx 0.72$, where the cross-polarized reflection vanishes, $r_{RL} = r_{RL} \approx 0$. 
At this point, the chiral super-crystal operates nearly as a handedness-preserving mirror \cite{Semnani2020, Voronin2022}. Near the band gap of the corresponding infinite structure, $\omega L/ (2 \pi c) \approx 0.44$ transmittance of RH light is completely suppressed, whereas LH light passes well through the structure.

\subsection{Helical As$_2$S$_3$ van der Waals homostructure as a chiral photonic super-crystal}

Practically, there is no need for an infinite number of twisted anisotropic layers to achieve configurational chirality. Our calculations show that even a few helically arranged anisotropic layers are sufficient to realize a chiral photonic super-crystal. To experimentally verify our computations, we used As$_2$S$_3$ van der Waals crystals since this material holds the record for the strong in-plane optical anisotropy in the visible range \cite{slavich2023exploring}. However, van der Waals layers of As$_2$S$_3$ are usually prepared by mechanical exfoliation \cite{Siskins2019} resulting in micrometer-scale samples, which complicates our test measurements \textit{via} traditional spectroscopic characterization techniques. Another complication is related to the exact manipulation of light's polarization required for the precise determination of optical chirality. On the contrary, our imaging spectroscopic ellipsometry technique resolves these issues, because it collects the signal with a lateral resolution down to 1 $\mu$m, and allows exact manipulation over the rotational degrees of the polarizer ($\varphi_P$), quarter-wave plate ($\varphi_C$), and the analyzer ($\varphi_A$). The schematic representation of our setup is shown in Figure \ref{fig3}(a). The Supporting Section S4 presents further details on it. Here, the super-crystal itself was assembled on As$_2$S$_3$ helical vdW homostructure in accordance with our theoretical evaluations. In particular, it consists of three layers of As$_2$S$_3$ with almost identical thicknesses ($\approx 160$ nm) twisted with respect to each other by a 30° misalignment angle in the case of each of the adjacent layers. Notably, the introduced twist degree accounts for the alignment of their crystallographic \textit{a}-axes (see Supporting Section S5 for the details on sample preparation). The optical micrograph of the assembled super-crystal based on As$_2$S$_3$ helical vdW homostructure is shown in Figure \ref{fig3}(b). Here, the dashed lines indicate the edges of the separate layers of As$_2$S$_3$. The "RoI" labels the relatively homogeneous region of interest with a diameter of $D \approx 10$ $\mu$m used in our measurements.

The explored dielectric response from separate As$_2$S$_3$ layers shows pronounced in-plane anisotropy, which is not very common for vdW crystals (see Figure \ref{fig3}(c)). It was measured \textit{via} polarized micro-transmission technique similar to the one reported in our previous work \cite{slavich2023exploring}. To demonstrate the effect of such in-plane anisotropy on our chiral photonic super-crystals, we investigated the transmission spectra for horizontal ($\widehat{H}$) and vertical ($\widehat{V}$) orientations of the polarizer, compensator, and the analyzer (see Figure \ref{fig3}(d, e)). While these measurements probe the in-plane anisotropic response of the structure rather than its chirality, a relatively good agreement between the experiment and theoretical calculations validates the accuracy of the determination of dielectric functions shown in Figure \ref{fig3}(c). 

To demonstrate the emergence of optical chirality, we further utilized imaging spectroscopic ellipsometry assuming the case of an ideal compensator and polarizer in our measurement setup (see Supporting Section S4 for details on the measurement technique). Here, a 45-degree rotation of the compensator ($\varphi_C$) was used to create right circular polarized (RCP) and left circularly polarized (LCP) lights passing through our super-crystal. 
Next, by subsequently setting the analyzer ($\varphi_A$) to horizontal and vertical positions we obtain $T(\widehat{L}$, $\widehat{V}$), $T(\widehat{R}$, $\widehat{V}$), $R(\widehat{L}$, $\widehat{H}$), $R(\widehat{R}$, $\widehat{H}$) signals as it is presented in Figure \ref{fig4}(a, b, c, e). It is worth mentioning that in this particular case, $T(\widehat{R}$, $\widehat{V}$) can be larger than 1 and $T(\widehat{R}$, $\widehat{H}$) can be smaller than 1, and \textit{vice versa}. The theoretical calculations are shown in Figure \ref{fig4}(a, b), and measurements are presented in Figure \ref{fig4}(d, e). In general, the transmission matrix of the sample can be written as
\begin{equation}
T_S = \left(\begin{array}{cc}
    t_{RR} & t_{RL}\\ 
    t_{LR} & t_{LL}
    \end{array}
    \right),
    \label{T-sample} 
\end{equation}
where $t_{RR}$, $t_{RL}$, $t_{LR}$, and $t_{LL}$ are the matrix elements of the full scattering matrix in the base of circular polarizations (see Supporting Section S3). Notably, measuring polarization-resolved intensity transmission amplitudes $T_R = |t_{RR}|^2$ and $T_L = |t_{LL}|^2$ is a complex task. However, evaluating the difference between $T_R$ and $T_L$ signals, which is of interest in the context of our chiral homostructures, is much more manageable when considering that $|t_{LR}|^2 = |t_{RL}|^2$ and measuring $T(\widehat{V}$) and $T(\widehat{H}$):
\begin{equation}
    T_R - T_L = \frac{T(\widehat{R}, \widehat{V}) + T(\widehat{R}, \widehat{H}) - T(\widehat{L}, \widehat{V}) - T(\widehat{L}, \widehat{H})}{2} .
    \label{dichroism}
\end{equation}
Our approximation holds rigorously only for systems possessing a horizontal $\pi$-rotation axis, which is rather violated in the case of the presence of the glass substrate. However, 
we demonstrate (see Supporting Section S4) that it leads to a negligible error of less than 4 $\%$ in the case of our helical homostructure. 
The corresponding differences between the $T_R$ - $T_L$ signals resulting in strong circular dichroism for both evaluated and measured cases are shown in Figure \ref{fig4}(c, f), respectively. Those are in good agreement considering the imperfections of the experimental setup. This circular dichroism is the best demonstration of the chiral nature of our photonic super-crystals based on As$_2$S$_3$ helical van der Waals homostructures.

\section*{Conclusion and Discussion}
To conclude, we have proposed and implemented experimentally a novel approach for engineering chiral photonic super-crystals based on helical van der Waals homostructures comprising biaxial anisotropic layers of an achiral nature. Our theoretical analysis demonstrates the emergence of a robust chiroptical response in super-crystals comprising a finite number of those layers. The effect was demonstrated experimentally with in-plane anisotropic As$_2$S$_3$ trilayers arranged in a chiral fashion into helical van der Waals homostructures with aligned crystallographic axes by means of a novel technique, which is based on imaging spectroscopic ellipsometry and allows direct detection of the circular dichroism – a straightforward consequence of an optical chirality. Our results pave the way to the creation of novel chiral systems composed of achiral crystals without the introduction of an additional patterning used in the creation of metasurfaces.





\bibliography{chirality}

\begin{thebibliography}{45}%
\makeatletter
\providecommand \@ifxundefined [1]{%
 \@ifx{#1\undefined}
}%
\providecommand \@ifnum [1]{%
 \ifnum #1\expandafter \@firstoftwo
 \else \expandafter \@secondoftwo
 \fi
}%
\providecommand \@ifx [1]{%
 \ifx #1\expandafter \@firstoftwo
 \else \expandafter \@secondoftwo
 \fi
}%
\providecommand \natexlab [1]{#1}%
\providecommand \enquote  [1]{``#1''}%
\providecommand \bibnamefont  [1]{#1}%
\providecommand \bibfnamefont [1]{#1}%
\providecommand \citenamefont [1]{#1}%
\providecommand \href@noop [0]{\@secondoftwo}%
\providecommand \href [0]{\begingroup \@sanitize@url \@href}%
\providecommand \@href[1]{\@@startlink{#1}\@@href}%
\providecommand \@@href[1]{\endgroup#1\@@endlink}%
\providecommand \@sanitize@url [0]{\catcode `\\12\catcode `\$12\catcode
  `\&12\catcode `\#12\catcode `\^12\catcode `\_12\catcode `\%12\relax}%
\providecommand \@@startlink[1]{}%
\providecommand \@@endlink[0]{}%
\providecommand \url  [0]{\begingroup\@sanitize@url \@url }%
\providecommand \@url [1]{\endgroup\@href {#1}{\urlprefix }}%
\providecommand \urlprefix  [0]{URL }%
\providecommand \Eprint [0]{\href }%
\providecommand \doibase [0]{https://doi.org/}%
\providecommand \selectlanguage [0]{\@gobble}%
\providecommand \bibinfo  [0]{\@secondoftwo}%
\providecommand \bibfield  [0]{\@secondoftwo}%
\providecommand \translation [1]{[#1]}%
\providecommand \BibitemOpen [0]{}%
\providecommand \bibitemStop [0]{}%
\providecommand \bibitemNoStop [0]{.\EOS\space}%
\providecommand \EOS [0]{\spacefactor3000\relax}%
\providecommand \BibitemShut  [1]{\csname bibitem#1\endcsname}%
\let\auto@bib@innerbib\@empty
\bibitem [{\citenamefont {Kelvin}(1894)}]{Kelvin1894}%
  \BibitemOpen
  \bibfield  {author} {\bibinfo {author} {\bibfnamefont {W.~T.~B.}\
  \bibnamefont {Kelvin}},\ }\href@noop {} {\emph {\bibinfo {title} {The
  Molecular Tactics of a Crystal}}},\ Robert Boyle lecture\ (\bibinfo
  {publisher} {Clarendon Press},\ \bibinfo {year} {1894})\BibitemShut {NoStop}%
\bibitem [{\citenamefont {Chhabra}\ \emph {et~al.}(2013)\citenamefont
  {Chhabra}, \citenamefont {Aseri},\ and\ \citenamefont
  {Padmanabhan}}]{chhabra2013review}%
  \BibitemOpen
  \bibfield  {author} {\bibinfo {author} {\bibfnamefont {N.}~\bibnamefont
  {Chhabra}}, \bibinfo {author} {\bibfnamefont {M.~L.}\ \bibnamefont {Aseri}},\
  and\ \bibinfo {author} {\bibfnamefont {D.}~\bibnamefont {Padmanabhan}},\
  }\bibfield  {title} {\bibinfo {title} {A review of drug isomerism and its
  significance},\ }\href@noop {} {\bibfield  {journal} {\bibinfo  {journal}
  {Int. J. Appl. Basic Med. Res.}\ }\textbf {\bibinfo {volume} {3}},\ \bibinfo
  {pages} {16} (\bibinfo {year} {2013})}\BibitemShut {NoStop}%
\bibitem [{\citenamefont {Asenjo-Garcia}\ and\ \citenamefont
  {De~Abajo}(2014)}]{asenjo2014dichroism}%
  \BibitemOpen
  \bibfield  {author} {\bibinfo {author} {\bibfnamefont {A.}~\bibnamefont
  {Asenjo-Garcia}}\ and\ \bibinfo {author} {\bibfnamefont {F.~G.}\ \bibnamefont
  {De~Abajo}},\ }\bibfield  {title} {\bibinfo {title} {Dichroism in the
  interaction between vortex electron beams, plasmons, and molecules},\
  }\href@noop {} {\bibfield  {journal} {\bibinfo  {journal} {Phys. Rev. Lett.}\
  }\textbf {\bibinfo {volume} {113}},\ \bibinfo {pages} {066102} (\bibinfo
  {year} {2014})}\BibitemShut {NoStop}%
\bibitem [{\citenamefont {Wallbank}\ \emph {et~al.}(2016)\citenamefont
  {Wallbank}, \citenamefont {Ghazaryan}, \citenamefont {Misra}, \citenamefont
  {Cao}, \citenamefont {Tu}, \citenamefont {Piot}, \citenamefont {Potemski},
  \citenamefont {Pezzini}, \citenamefont {Wiedmann}, \citenamefont {Zeitler},
  \citenamefont {Lane}, \citenamefont {Morozov}, \citenamefont {Greenaway},
  \citenamefont {Eaves}, \citenamefont {Geim}, \citenamefont
  {Fal{\textquotesingle}ko}, \citenamefont {Novoselov},\ and\ \citenamefont
  {Mishchenko}}]{Wallbank2016}%
  \BibitemOpen
  \bibfield  {author} {\bibinfo {author} {\bibfnamefont {J.~R.}\ \bibnamefont
  {Wallbank}}, \bibinfo {author} {\bibfnamefont {D.}~\bibnamefont {Ghazaryan}},
  \bibinfo {author} {\bibfnamefont {A.}~\bibnamefont {Misra}}, \bibinfo
  {author} {\bibfnamefont {Y.}~\bibnamefont {Cao}}, \bibinfo {author}
  {\bibfnamefont {J.~S.}\ \bibnamefont {Tu}}, \bibinfo {author} {\bibfnamefont
  {B.~A.}\ \bibnamefont {Piot}}, \bibinfo {author} {\bibfnamefont
  {M.}~\bibnamefont {Potemski}}, \bibinfo {author} {\bibfnamefont
  {S.}~\bibnamefont {Pezzini}}, \bibinfo {author} {\bibfnamefont
  {S.}~\bibnamefont {Wiedmann}}, \bibinfo {author} {\bibfnamefont
  {U.}~\bibnamefont {Zeitler}}, \bibinfo {author} {\bibfnamefont {T.~L.~M.}\
  \bibnamefont {Lane}}, \bibinfo {author} {\bibfnamefont {S.~V.}\ \bibnamefont
  {Morozov}}, \bibinfo {author} {\bibfnamefont {M.~T.}\ \bibnamefont
  {Greenaway}}, \bibinfo {author} {\bibfnamefont {L.}~\bibnamefont {Eaves}},
  \bibinfo {author} {\bibfnamefont {A.~K.}\ \bibnamefont {Geim}}, \bibinfo
  {author} {\bibfnamefont {V.~I.}\ \bibnamefont {Fal{\textquotesingle}ko}},
  \bibinfo {author} {\bibfnamefont {K.~S.}\ \bibnamefont {Novoselov}},\ and\
  \bibinfo {author} {\bibfnamefont {A.}~\bibnamefont {Mishchenko}},\ }\bibfield
   {title} {\bibinfo {title} {Tuning the valley and chiral quantum state of
  dirac electrons in van der waals heterostructures},\ }\href
  {https://doi.org/10.1126/science.aaf4621} {\bibfield  {journal} {\bibinfo
  {journal} {Science}\ }\textbf {\bibinfo {volume} {353}},\ \bibinfo {pages}
  {575} (\bibinfo {year} {2016})}\BibitemShut {NoStop}%
\bibitem [{\citenamefont {Stickler}\ \emph {et~al.}(2021)\citenamefont
  {Stickler}, \citenamefont {Diekmann}, \citenamefont {Berger},\ and\
  \citenamefont {Wang}}]{Stickler2021}%
  \BibitemOpen
  \bibfield  {author} {\bibinfo {author} {\bibfnamefont {B.~A.}\ \bibnamefont
  {Stickler}}, \bibinfo {author} {\bibfnamefont {M.}~\bibnamefont {Diekmann}},
  \bibinfo {author} {\bibfnamefont {R.}~\bibnamefont {Berger}},\ and\ \bibinfo
  {author} {\bibfnamefont {D.}~\bibnamefont {Wang}},\ }\bibfield  {title}
  {\bibinfo {title} {{Enantiomer Superpositions from Matter-Wave Interference
  of Chiral Molecules}},\ }\href@noop {} {\bibfield  {journal} {\bibinfo
  {journal} {Phys. Rev. X}\ }\textbf {\bibinfo {volume} {11}},\ \bibinfo
  {pages} {31056} (\bibinfo {year} {2021})}\BibitemShut {NoStop}%
\bibitem [{\citenamefont {Lindell}\ \emph {et~al.}(2018)\citenamefont
  {Lindell}, \citenamefont {Sihvola}, \citenamefont {Tretyakov},\ and\
  \citenamefont {Vitanen}}]{Lindell2018}%
  \BibitemOpen
  \bibfield  {author} {\bibinfo {author} {\bibfnamefont {I.}~\bibnamefont
  {Lindell}}, \bibinfo {author} {\bibfnamefont {A.}~\bibnamefont {Sihvola}},
  \bibinfo {author} {\bibfnamefont {S.}~\bibnamefont {Tretyakov}},\ and\
  \bibinfo {author} {\bibfnamefont {A.}~\bibnamefont {Vitanen}},\ }\href@noop
  {} {\emph {\bibinfo {title} {{Electromagnetic waves in chiral and
  Bi-isotropic media}}}}\ (\bibinfo  {publisher} {Artech House},\ \bibinfo
  {year} {2018})\ p.\ \bibinfo {pages} {332}\BibitemShut {NoStop}%
\bibitem [{\citenamefont {Inoue}\ and\ \citenamefont
  {Ramamurthy}(2004)}]{Inoue2004}%
  \BibitemOpen
  \bibfield  {author} {\bibinfo {author} {\bibfnamefont {Y.}~\bibnamefont
  {Inoue}}\ and\ \bibinfo {author} {\bibfnamefont {V.}~\bibnamefont
  {Ramamurthy}},\ }\href@noop {} {\emph {\bibinfo {title} {{Chiral
  photochemistry}}}}\ (\bibinfo  {publisher} {Marcel Dekker},\ \bibinfo {year}
  {2004})\ p.\ \bibinfo {pages} {685}\BibitemShut {NoStop}%
\bibitem [{\citenamefont {Govorov}\ \emph {et~al.}(2010)\citenamefont
  {Govorov}, \citenamefont {Fan}, \citenamefont {Hernandez}, \citenamefont
  {Slocik},\ and\ \citenamefont {Naik}}]{Govorov2010}%
  \BibitemOpen
  \bibfield  {author} {\bibinfo {author} {\bibfnamefont {A.~O.}\ \bibnamefont
  {Govorov}}, \bibinfo {author} {\bibfnamefont {Z.}~\bibnamefont {Fan}},
  \bibinfo {author} {\bibfnamefont {P.}~\bibnamefont {Hernandez}}, \bibinfo
  {author} {\bibfnamefont {J.~M.}\ \bibnamefont {Slocik}},\ and\ \bibinfo
  {author} {\bibfnamefont {R.~R.}\ \bibnamefont {Naik}},\ }\bibfield  {title}
  {\bibinfo {title} {{Theory of circular dichroism of nanomaterials comprising
  chiral molecules and nanocrystals: Plasmon enhancement, dipole interactions,
  and dielectric effects}},\ }\href {https://doi.org/10.1021/nl100010v}
  {\bibfield  {journal} {\bibinfo  {journal} {Nano Lett.}\ }\textbf {\bibinfo
  {volume} {10}},\ \bibinfo {pages} {1374} (\bibinfo {year}
  {2010})}\BibitemShut {NoStop}%
\bibitem [{\citenamefont {Tang}\ and\ \citenamefont {Cohen}(2010)}]{Tang2010}%
  \BibitemOpen
  \bibfield  {author} {\bibinfo {author} {\bibfnamefont {Y.}~\bibnamefont
  {Tang}}\ and\ \bibinfo {author} {\bibfnamefont {A.~E.}\ \bibnamefont
  {Cohen}},\ }\bibfield  {title} {\bibinfo {title} {{Optical chirality and its
  interaction with matter}},\ }\href
  {https://doi.org/10.1103/PhysRevLett.104.163901} {\bibfield  {journal}
  {\bibinfo  {journal} {Phys. Rev. Lett.}\ }\textbf {\bibinfo {volume} {104}},\
  \bibinfo {pages} {163901} (\bibinfo {year} {2010})}\BibitemShut {NoStop}%
\bibitem [{\citenamefont {Hendry}\ \emph {et~al.}(2010)\citenamefont {Hendry},
  \citenamefont {Carpy}, \citenamefont {Johnston}, \citenamefont {Popland},
  \citenamefont {Mikhaylovskiy}, \citenamefont {Lapthorn}, \citenamefont
  {Kelly}, \citenamefont {Barron}, \citenamefont {Gadegaard},\ and\
  \citenamefont {Kadodwala}}]{Hendry2010}%
  \BibitemOpen
  \bibfield  {author} {\bibinfo {author} {\bibfnamefont {E.}~\bibnamefont
  {Hendry}}, \bibinfo {author} {\bibfnamefont {T.}~\bibnamefont {Carpy}},
  \bibinfo {author} {\bibfnamefont {J.}~\bibnamefont {Johnston}}, \bibinfo
  {author} {\bibfnamefont {M.}~\bibnamefont {Popland}}, \bibinfo {author}
  {\bibfnamefont {R.~V.}\ \bibnamefont {Mikhaylovskiy}}, \bibinfo {author}
  {\bibfnamefont {A.~J.}\ \bibnamefont {Lapthorn}}, \bibinfo {author}
  {\bibfnamefont {S.~M.}\ \bibnamefont {Kelly}}, \bibinfo {author}
  {\bibfnamefont {L.~D.}\ \bibnamefont {Barron}}, \bibinfo {author}
  {\bibfnamefont {N.}~\bibnamefont {Gadegaard}},\ and\ \bibinfo {author}
  {\bibfnamefont {M.}~\bibnamefont {Kadodwala}},\ }\bibfield  {title} {\bibinfo
  {title} {{Ultrasensitive detection and characterization of biomolecules using
  superchiral fields}},\ }\href {https://doi.org/10.1038/nnano.2010.209}
  {\bibfield  {journal} {\bibinfo  {journal} {Nat. Nanotechnol.}\ }\textbf
  {\bibinfo {volume} {5}},\ \bibinfo {pages} {783} (\bibinfo {year}
  {2010})}\BibitemShut {NoStop}%
\bibitem [{\citenamefont {Gorkunov}\ \emph {et~al.}(2020)\citenamefont
  {Gorkunov}, \citenamefont {Antonov},\ and\ \citenamefont
  {Kivshar}}]{Gorkunov2020}%
  \BibitemOpen
  \bibfield  {author} {\bibinfo {author} {\bibfnamefont {M.~V.}\ \bibnamefont
  {Gorkunov}}, \bibinfo {author} {\bibfnamefont {A.~A.}\ \bibnamefont
  {Antonov}},\ and\ \bibinfo {author} {\bibfnamefont {Y.~S.}\ \bibnamefont
  {Kivshar}},\ }\bibfield  {title} {\bibinfo {title} {{Metasurfaces with
  Maximum Chirality Empowered by Bound States in the Continuum}},\ }\href
  {https://doi.org/10.1103/PhysRevLett.125.093903} {\bibfield  {journal}
  {\bibinfo  {journal} {Phys. Rev. Lett.}\ }\textbf {\bibinfo {volume} {125}},\
  \bibinfo {pages} {93903} (\bibinfo {year} {2020})}\BibitemShut {NoStop}%
\bibitem [{\citenamefont {Chen}\ \emph {et~al.}(2023)\citenamefont {Chen},
  \citenamefont {Deng}, \citenamefont {Sha}, \citenamefont {Chen},
  \citenamefont {Wang}, \citenamefont {Chen}, \citenamefont {Wu}, \citenamefont
  {Chu}, \citenamefont {Kivshar}, \citenamefont {Xiao} \emph
  {et~al.}}]{chen2023observation}%
  \BibitemOpen
  \bibfield  {author} {\bibinfo {author} {\bibfnamefont {Y.}~\bibnamefont
  {Chen}}, \bibinfo {author} {\bibfnamefont {H.}~\bibnamefont {Deng}}, \bibinfo
  {author} {\bibfnamefont {X.}~\bibnamefont {Sha}}, \bibinfo {author}
  {\bibfnamefont {W.}~\bibnamefont {Chen}}, \bibinfo {author} {\bibfnamefont
  {R.}~\bibnamefont {Wang}}, \bibinfo {author} {\bibfnamefont {Y.-H.}\
  \bibnamefont {Chen}}, \bibinfo {author} {\bibfnamefont {D.}~\bibnamefont
  {Wu}}, \bibinfo {author} {\bibfnamefont {J.}~\bibnamefont {Chu}}, \bibinfo
  {author} {\bibfnamefont {Y.~S.}\ \bibnamefont {Kivshar}}, \bibinfo {author}
  {\bibfnamefont {S.}~\bibnamefont {Xiao}}, \emph {et~al.},\ }\bibfield
  {title} {\bibinfo {title} {Observation of intrinsic chiral bound states in
  the continuum},\ }\href@noop {} {\bibfield  {journal} {\bibinfo  {journal}
  {Nature}\ }\textbf {\bibinfo {volume} {613}},\ \bibinfo {pages} {474}
  (\bibinfo {year} {2023})}\BibitemShut {NoStop}%
\bibitem [{\citenamefont {Plum}\ and\ \citenamefont
  {Zheludev}(2015)}]{plum2015chiral}%
  \BibitemOpen
  \bibfield  {author} {\bibinfo {author} {\bibfnamefont {E.}~\bibnamefont
  {Plum}}\ and\ \bibinfo {author} {\bibfnamefont {N.~I.}\ \bibnamefont
  {Zheludev}},\ }\bibfield  {title} {\bibinfo {title} {Chiral mirrors},\
  }\href@noop {} {\bibfield  {journal} {\bibinfo  {journal} {Applied Physics
  Letters}\ }\textbf {\bibinfo {volume} {106}},\ \bibinfo {pages} {221901}
  (\bibinfo {year} {2015})}\BibitemShut {NoStop}%
\bibitem [{\citenamefont {Feis}\ \emph {et~al.}(2020)\citenamefont {Feis},
  \citenamefont {Beutel}, \citenamefont {K{\"{o}}pfler}, \citenamefont
  {Garcia-Santiago}, \citenamefont {Rockstuhl}, \citenamefont {Wegener},\ and\
  \citenamefont {Fernandez-Corbaton}}]{Feis2020}%
  \BibitemOpen
  \bibfield  {author} {\bibinfo {author} {\bibfnamefont {J.}~\bibnamefont
  {Feis}}, \bibinfo {author} {\bibfnamefont {D.}~\bibnamefont {Beutel}},
  \bibinfo {author} {\bibfnamefont {J.}~\bibnamefont {K{\"{o}}pfler}}, \bibinfo
  {author} {\bibfnamefont {X.}~\bibnamefont {Garcia-Santiago}}, \bibinfo
  {author} {\bibfnamefont {C.}~\bibnamefont {Rockstuhl}}, \bibinfo {author}
  {\bibfnamefont {M.}~\bibnamefont {Wegener}},\ and\ \bibinfo {author}
  {\bibfnamefont {I.}~\bibnamefont {Fernandez-Corbaton}},\ }\bibfield  {title}
  {\bibinfo {title} {{Helicity-Preserving Optical Cavity Modes for Enhanced
  Sensing of Chiral Molecules}},\ }\href
  {https://doi.org/10.1103/PhysRevLett.124.033201} {\bibfield  {journal}
  {\bibinfo  {journal} {Phys. Rev. Lett.}\ }\textbf {\bibinfo {volume} {124}},\
  \bibinfo {pages} {033201} (\bibinfo {year} {2020})}\BibitemShut {NoStop}%
\bibitem [{\citenamefont {Voronin}\ \emph {et~al.}(2022)\citenamefont
  {Voronin}, \citenamefont {Taradin}, \citenamefont {Gorkunov},\ and\
  \citenamefont {Baranov}}]{Voronin2022}%
  \BibitemOpen
  \bibfield  {author} {\bibinfo {author} {\bibfnamefont {K.}~\bibnamefont
  {Voronin}}, \bibinfo {author} {\bibfnamefont {A.~S.}\ \bibnamefont
  {Taradin}}, \bibinfo {author} {\bibfnamefont {M.~V.}\ \bibnamefont
  {Gorkunov}},\ and\ \bibinfo {author} {\bibfnamefont {D.~G.}\ \bibnamefont
  {Baranov}},\ }\bibfield  {title} {\bibinfo {title} {Single-handedness chiral
  optical cavities},\ }\href@noop {} {\bibfield  {journal} {\bibinfo  {journal}
  {ACS Photonics}\ }\textbf {\bibinfo {volume} {9}},\ \bibinfo {pages} {2652}
  (\bibinfo {year} {2022})}\BibitemShut {NoStop}%
\bibitem [{\citenamefont {Schafer}\ and\ \citenamefont
  {Baranov}(2023)}]{schafer2023chiral}%
  \BibitemOpen
  \bibfield  {author} {\bibinfo {author} {\bibfnamefont {C.}~\bibnamefont
  {Schafer}}\ and\ \bibinfo {author} {\bibfnamefont {D.~G.}\ \bibnamefont
  {Baranov}},\ }\bibfield  {title} {\bibinfo {title} {Chiral polaritonics:
  Analytical solutions, intuition, and use},\ }\href@noop {} {\bibfield
  {journal} {\bibinfo  {journal} {J. Phys. Chem. C}\ }\textbf {\bibinfo
  {volume} {14}},\ \bibinfo {pages} {3777} (\bibinfo {year}
  {2023})}\BibitemShut {NoStop}%
\bibitem [{\citenamefont {Mauro}\ \emph {et~al.}(2023)\citenamefont {Mauro},
  \citenamefont {Fregoni}, \citenamefont {Feist},\ and\ \citenamefont
  {Avriller}}]{PhysRevA.107.L021501}%
  \BibitemOpen
  \bibfield  {author} {\bibinfo {author} {\bibfnamefont {L.}~\bibnamefont
  {Mauro}}, \bibinfo {author} {\bibfnamefont {J.}~\bibnamefont {Fregoni}},
  \bibinfo {author} {\bibfnamefont {J.}~\bibnamefont {Feist}},\ and\ \bibinfo
  {author} {\bibfnamefont {R.}~\bibnamefont {Avriller}},\ }\bibfield  {title}
  {\bibinfo {title} {Chiral discrimination in helicity-preserving fabry-p\'erot
  cavities},\ }\href {https://doi.org/10.1103/PhysRevA.107.L021501} {\bibfield
  {journal} {\bibinfo  {journal} {Phys. Rev. A}\ }\textbf {\bibinfo {volume}
  {107}},\ \bibinfo {pages} {L021501} (\bibinfo {year} {2023})}\BibitemShut
  {NoStop}%
\bibitem [{\citenamefont {Han}\ \emph {et~al.}(2023)\citenamefont {Han},
  \citenamefont {Wang}, \citenamefont {Sun}, \citenamefont {Wang},\ and\
  \citenamefont {Tang}}]{han2023recent}%
  \BibitemOpen
  \bibfield  {author} {\bibinfo {author} {\bibfnamefont {Z.}~\bibnamefont
  {Han}}, \bibinfo {author} {\bibfnamefont {F.}~\bibnamefont {Wang}}, \bibinfo
  {author} {\bibfnamefont {J.}~\bibnamefont {Sun}}, \bibinfo {author}
  {\bibfnamefont {X.}~\bibnamefont {Wang}},\ and\ \bibinfo {author}
  {\bibfnamefont {Z.}~\bibnamefont {Tang}},\ }\bibfield  {title} {\bibinfo
  {title} {Recent advances in ultrathin chiral metasurfaces by twisted
  stacking},\ }\href@noop {} {\bibfield  {journal} {\bibinfo  {journal} {Adv.
  Mater.}\ }\textbf {\bibinfo {volume} {35}},\ \bibinfo {pages} {2206141}
  (\bibinfo {year} {2023})}\BibitemShut {NoStop}%
\bibitem [{\citenamefont {Hentschel}\ \emph {et~al.}(2012)\citenamefont
  {Hentschel}, \citenamefont {Sch{\"a}ferling}, \citenamefont {Weiss},
  \citenamefont {Liu},\ and\ \citenamefont {Giessen}}]{hentschel2012three}%
  \BibitemOpen
  \bibfield  {author} {\bibinfo {author} {\bibfnamefont {M.}~\bibnamefont
  {Hentschel}}, \bibinfo {author} {\bibfnamefont {M.}~\bibnamefont
  {Sch{\"a}ferling}}, \bibinfo {author} {\bibfnamefont {T.}~\bibnamefont
  {Weiss}}, \bibinfo {author} {\bibfnamefont {N.}~\bibnamefont {Liu}},\ and\
  \bibinfo {author} {\bibfnamefont {H.}~\bibnamefont {Giessen}},\ }\bibfield
  {title} {\bibinfo {title} {Three-dimensional chiral plasmonic oligomers},\
  }\href@noop {} {\bibfield  {journal} {\bibinfo  {journal} {Nano Lett.}\
  }\textbf {\bibinfo {volume} {12}},\ \bibinfo {pages} {2542} (\bibinfo {year}
  {2012})}\BibitemShut {NoStop}%
\bibitem [{\citenamefont {Sch{\"a}ferling}\ \emph {et~al.}(2012)\citenamefont
  {Sch{\"a}ferling}, \citenamefont {Dregely}, \citenamefont {Hentschel},\ and\
  \citenamefont {Giessen}}]{schaferling2012tailoring}%
  \BibitemOpen
  \bibfield  {author} {\bibinfo {author} {\bibfnamefont {M.}~\bibnamefont
  {Sch{\"a}ferling}}, \bibinfo {author} {\bibfnamefont {D.}~\bibnamefont
  {Dregely}}, \bibinfo {author} {\bibfnamefont {M.}~\bibnamefont {Hentschel}},\
  and\ \bibinfo {author} {\bibfnamefont {H.}~\bibnamefont {Giessen}},\
  }\bibfield  {title} {\bibinfo {title} {Tailoring enhanced optical chirality:
  design principles for chiral plasmonic nanostructures},\ }\href@noop {}
  {\bibfield  {journal} {\bibinfo  {journal} {Phys. Rev. X}\ }\textbf {\bibinfo
  {volume} {2}},\ \bibinfo {pages} {031010} (\bibinfo {year}
  {2012})}\BibitemShut {NoStop}%
\bibitem [{\citenamefont {Avalos-Ovando}\ \emph {et~al.}(2022)\citenamefont
  {Avalos-Ovando}, \citenamefont {Santiago}, \citenamefont {Movsesyan},
  \citenamefont {Kong}, \citenamefont {Yu}, \citenamefont {Besteiro},
  \citenamefont {Khorashad}, \citenamefont {Okamoto}, \citenamefont {Slocik},
  \citenamefont {Correa-Duarte} \emph {et~al.}}]{avalos2022chiral}%
  \BibitemOpen
  \bibfield  {author} {\bibinfo {author} {\bibfnamefont {O.}~\bibnamefont
  {Avalos-Ovando}}, \bibinfo {author} {\bibfnamefont {E.~Y.}\ \bibnamefont
  {Santiago}}, \bibinfo {author} {\bibfnamefont {A.}~\bibnamefont {Movsesyan}},
  \bibinfo {author} {\bibfnamefont {X.-T.}\ \bibnamefont {Kong}}, \bibinfo
  {author} {\bibfnamefont {P.}~\bibnamefont {Yu}}, \bibinfo {author}
  {\bibfnamefont {L.~V.}\ \bibnamefont {Besteiro}}, \bibinfo {author}
  {\bibfnamefont {L.~K.}\ \bibnamefont {Khorashad}}, \bibinfo {author}
  {\bibfnamefont {H.}~\bibnamefont {Okamoto}}, \bibinfo {author} {\bibfnamefont
  {J.~M.}\ \bibnamefont {Slocik}}, \bibinfo {author} {\bibfnamefont {M.~A.}\
  \bibnamefont {Correa-Duarte}}, \emph {et~al.},\ }\bibfield  {title} {\bibinfo
  {title} {Chiral bioinspired plasmonics: A paradigm shift for optical activity
  and photochemistry},\ }\href@noop {} {\bibfield  {journal} {\bibinfo
  {journal} {ACS Photonics}\ } (\bibinfo {year} {2022})}\BibitemShut {NoStop}%
\bibitem [{\citenamefont {Cecconello}\ \emph {et~al.}(2017)\citenamefont
  {Cecconello}, \citenamefont {Besteiro}, \citenamefont {Govorov},\ and\
  \citenamefont {Willner}}]{Cecconello2017}%
  \BibitemOpen
  \bibfield  {author} {\bibinfo {author} {\bibfnamefont {A.}~\bibnamefont
  {Cecconello}}, \bibinfo {author} {\bibfnamefont {L.~V.}\ \bibnamefont
  {Besteiro}}, \bibinfo {author} {\bibfnamefont {A.~O.}\ \bibnamefont
  {Govorov}},\ and\ \bibinfo {author} {\bibfnamefont {I.}~\bibnamefont
  {Willner}},\ }\bibfield  {title} {\bibinfo {title} {{Chiroplasmonic DNA-based
  nanostructures}},\ }\href {https://doi.org/10.1038/natrevmats.2017.39}
  {\bibfield  {journal} {\bibinfo  {journal} {Nat. Rev. Mater.}\ }\textbf
  {\bibinfo {volume} {2}},\ \bibinfo {pages} {17039} (\bibinfo {year}
  {2017})}\BibitemShut {NoStop}%
\bibitem [{\citenamefont {Kim}\ \emph {et~al.}(2016)\citenamefont {Kim},
  \citenamefont {S{\'a}nchez-Castillo}, \citenamefont {Ziegler}, \citenamefont
  {Ogawa}, \citenamefont {Noguez},\ and\ \citenamefont {Park}}]{kim2016chiral}%
  \BibitemOpen
  \bibfield  {author} {\bibinfo {author} {\bibfnamefont {C.-J.}\ \bibnamefont
  {Kim}}, \bibinfo {author} {\bibfnamefont {A.}~\bibnamefont
  {S{\'a}nchez-Castillo}}, \bibinfo {author} {\bibfnamefont {Z.}~\bibnamefont
  {Ziegler}}, \bibinfo {author} {\bibfnamefont {Y.}~\bibnamefont {Ogawa}},
  \bibinfo {author} {\bibfnamefont {C.}~\bibnamefont {Noguez}},\ and\ \bibinfo
  {author} {\bibfnamefont {J.}~\bibnamefont {Park}},\ }\bibfield  {title}
  {\bibinfo {title} {Chiral atomically thin films},\ }\href@noop {} {\bibfield
  {journal} {\bibinfo  {journal} {Nat. Nanotechnol.}\ }\textbf {\bibinfo
  {volume} {11}},\ \bibinfo {pages} {520} (\bibinfo {year} {2016})}\BibitemShut
  {NoStop}%
\bibitem [{\citenamefont {Ochoa}\ and\ \citenamefont
  {Asenjo-Garcia}(2020)}]{ochoa2020flat}%
  \BibitemOpen
  \bibfield  {author} {\bibinfo {author} {\bibfnamefont {H.}~\bibnamefont
  {Ochoa}}\ and\ \bibinfo {author} {\bibfnamefont {A.}~\bibnamefont
  {Asenjo-Garcia}},\ }\bibfield  {title} {\bibinfo {title} {Flat bands and
  chiral optical response of moir{\'e} insulators},\ }\href@noop {} {\bibfield
  {journal} {\bibinfo  {journal} {Phys. Rev. Lett.}\ }\textbf {\bibinfo
  {volume} {125}},\ \bibinfo {pages} {037402} (\bibinfo {year}
  {2020})}\BibitemShut {NoStop}%
\bibitem [{\citenamefont {Wu}\ \emph {et~al.}(2022)\citenamefont {Wu},
  \citenamefont {Zheng}, \citenamefont {Li}, \citenamefont {Ding},
  \citenamefont {He}, \citenamefont {Zeng}, \citenamefont {Chen}, \citenamefont
  {Liu}, \citenamefont {Chen}, \citenamefont {Pan} \emph {et~al.}}]{Wu2022}%
  \BibitemOpen
  \bibfield  {author} {\bibinfo {author} {\bibfnamefont {B.}~\bibnamefont
  {Wu}}, \bibinfo {author} {\bibfnamefont {H.}~\bibnamefont {Zheng}}, \bibinfo
  {author} {\bibfnamefont {S.}~\bibnamefont {Li}}, \bibinfo {author}
  {\bibfnamefont {J.}~\bibnamefont {Ding}}, \bibinfo {author} {\bibfnamefont
  {J.}~\bibnamefont {He}}, \bibinfo {author} {\bibfnamefont {Y.}~\bibnamefont
  {Zeng}}, \bibinfo {author} {\bibfnamefont {K.}~\bibnamefont {Chen}}, \bibinfo
  {author} {\bibfnamefont {Z.}~\bibnamefont {Liu}}, \bibinfo {author}
  {\bibfnamefont {S.}~\bibnamefont {Chen}}, \bibinfo {author} {\bibfnamefont
  {A.}~\bibnamefont {Pan}}, \emph {et~al.},\ }\bibfield  {title} {\bibinfo
  {title} {Evidence for moir{\'e} intralayer excitons in twisted wse2/wse2
  homobilayer superlattices},\ }\href@noop {} {\bibfield  {journal} {\bibinfo
  {journal} {Light Sci. Appl.}\ }\textbf {\bibinfo {volume} {11}},\ \bibinfo
  {pages} {166} (\bibinfo {year} {2022})}\BibitemShut {NoStop}%
\bibitem [{\citenamefont {F{\"o}rg}\ \emph {et~al.}(2021)\citenamefont
  {F{\"o}rg}, \citenamefont {Baimuratov}, \citenamefont {Kruchinin},
  \citenamefont {Vovk}, \citenamefont {Scherzer}, \citenamefont {F{\"o}rste},
  \citenamefont {Funk}, \citenamefont {Watanabe}, \citenamefont {Taniguchi},\
  and\ \citenamefont {H{\"o}gele}}]{forg2021moire}%
  \BibitemOpen
  \bibfield  {author} {\bibinfo {author} {\bibfnamefont {M.}~\bibnamefont
  {F{\"o}rg}}, \bibinfo {author} {\bibfnamefont {A.~S.}\ \bibnamefont
  {Baimuratov}}, \bibinfo {author} {\bibfnamefont {S.~Y.}\ \bibnamefont
  {Kruchinin}}, \bibinfo {author} {\bibfnamefont {I.~A.}\ \bibnamefont {Vovk}},
  \bibinfo {author} {\bibfnamefont {J.}~\bibnamefont {Scherzer}}, \bibinfo
  {author} {\bibfnamefont {J.}~\bibnamefont {F{\"o}rste}}, \bibinfo {author}
  {\bibfnamefont {V.}~\bibnamefont {Funk}}, \bibinfo {author} {\bibfnamefont
  {K.}~\bibnamefont {Watanabe}}, \bibinfo {author} {\bibfnamefont
  {T.}~\bibnamefont {Taniguchi}},\ and\ \bibinfo {author} {\bibfnamefont
  {A.}~\bibnamefont {H{\"o}gele}},\ }\bibfield  {title} {\bibinfo {title}
  {Moir{\'e} excitons in mose2-wse2 heterobilayers and heterotrilayers},\
  }\href@noop {} {\bibfield  {journal} {\bibinfo  {journal} {Nat. Commun.}\
  }\textbf {\bibinfo {volume} {12}},\ \bibinfo {pages} {1656} (\bibinfo {year}
  {2021})}\BibitemShut {NoStop}%
\bibitem [{\citenamefont {Chen}\ \emph {et~al.}(2022)\citenamefont {Chen},
  \citenamefont {Lian}, \citenamefont {Huang}, \citenamefont {Su},
  \citenamefont {Rashetnia}, \citenamefont {Yan}, \citenamefont {Blei},
  \citenamefont {Taniguchi}, \citenamefont {Watanabe}, \citenamefont {Tongay}
  \emph {et~al.}}]{Chen2022}%
  \BibitemOpen
  \bibfield  {author} {\bibinfo {author} {\bibfnamefont {D.}~\bibnamefont
  {Chen}}, \bibinfo {author} {\bibfnamefont {Z.}~\bibnamefont {Lian}}, \bibinfo
  {author} {\bibfnamefont {X.}~\bibnamefont {Huang}}, \bibinfo {author}
  {\bibfnamefont {Y.}~\bibnamefont {Su}}, \bibinfo {author} {\bibfnamefont
  {M.}~\bibnamefont {Rashetnia}}, \bibinfo {author} {\bibfnamefont
  {L.}~\bibnamefont {Yan}}, \bibinfo {author} {\bibfnamefont {M.}~\bibnamefont
  {Blei}}, \bibinfo {author} {\bibfnamefont {T.}~\bibnamefont {Taniguchi}},
  \bibinfo {author} {\bibfnamefont {K.}~\bibnamefont {Watanabe}}, \bibinfo
  {author} {\bibfnamefont {S.}~\bibnamefont {Tongay}}, \emph {et~al.},\
  }\bibfield  {title} {\bibinfo {title} {Tuning moir{\'e} excitons and
  correlated electronic states through layer degree of freedom},\ }\href
  {https://doi.org/10.1038/s41467-022-32493-9} {\bibfield  {journal} {\bibinfo
  {journal} {Nat. Commun.}\ }\textbf {\bibinfo {volume} {13}},\ \bibinfo
  {pages} {4810} (\bibinfo {year} {2022})}\BibitemShut {NoStop}%
\bibitem [{\citenamefont {Li}\ \emph {et~al.}(2014)\citenamefont {Li},
  \citenamefont {Chernikov}, \citenamefont {Zhang}, \citenamefont {Rigosi},
  \citenamefont {Hill}, \citenamefont {Van Der~Zande}, \citenamefont {Chenet},
  \citenamefont {Shih}, \citenamefont {Hone},\ and\ \citenamefont
  {Heinz}}]{li2014measurement}%
  \BibitemOpen
  \bibfield  {author} {\bibinfo {author} {\bibfnamefont {Y.}~\bibnamefont
  {Li}}, \bibinfo {author} {\bibfnamefont {A.}~\bibnamefont {Chernikov}},
  \bibinfo {author} {\bibfnamefont {X.}~\bibnamefont {Zhang}}, \bibinfo
  {author} {\bibfnamefont {A.}~\bibnamefont {Rigosi}}, \bibinfo {author}
  {\bibfnamefont {H.~M.}\ \bibnamefont {Hill}}, \bibinfo {author}
  {\bibfnamefont {A.~M.}\ \bibnamefont {Van Der~Zande}}, \bibinfo {author}
  {\bibfnamefont {D.~A.}\ \bibnamefont {Chenet}}, \bibinfo {author}
  {\bibfnamefont {E.-M.}\ \bibnamefont {Shih}}, \bibinfo {author}
  {\bibfnamefont {J.}~\bibnamefont {Hone}},\ and\ \bibinfo {author}
  {\bibfnamefont {T.~F.}\ \bibnamefont {Heinz}},\ }\bibfield  {title} {\bibinfo
  {title} {Measurement of the optical dielectric function of monolayer
  transition-metal dichalcogenides: ${\mathrm{mos}}_{2}$,
  $\mathrm{Mo}\mathrm{S}{\mathrm{e}}_{2}$, ${\mathrm{ws}}_{2}$, and
  $\mathrm{WS}{\mathrm{e}}_{2}$},\ }\href@noop {} {\bibfield  {journal}
  {\bibinfo  {journal} {Physical Review B}\ }\textbf {\bibinfo {volume} {90}},\
  \bibinfo {pages} {205422} (\bibinfo {year} {2014})}\BibitemShut {NoStop}%
\bibitem [{\citenamefont {Mak}\ and\ \citenamefont
  {Shan}(2016)}]{mak2016photonics}%
  \BibitemOpen
  \bibfield  {author} {\bibinfo {author} {\bibfnamefont {K.~F.}\ \bibnamefont
  {Mak}}\ and\ \bibinfo {author} {\bibfnamefont {J.}~\bibnamefont {Shan}},\
  }\bibfield  {title} {\bibinfo {title} {Photonics and optoelectronics of 2d
  semiconductor transition metal dichalcogenides},\ }\href@noop {} {\bibfield
  {journal} {\bibinfo  {journal} {Nature Photonics}\ }\textbf {\bibinfo
  {volume} {10}},\ \bibinfo {pages} {216} (\bibinfo {year} {2016})}\BibitemShut
  {NoStop}%
\bibitem [{\citenamefont {Khaliji}\ \emph {et~al.}(2022)\citenamefont
  {Khaliji}, \citenamefont {Mart{\'\i}n-Moreno}, \citenamefont {Avouris},
  \citenamefont {Oh},\ and\ \citenamefont {Low}}]{khaliji2022twisted}%
  \BibitemOpen
  \bibfield  {author} {\bibinfo {author} {\bibfnamefont {K.}~\bibnamefont
  {Khaliji}}, \bibinfo {author} {\bibfnamefont {L.}~\bibnamefont
  {Mart{\'\i}n-Moreno}}, \bibinfo {author} {\bibfnamefont {P.}~\bibnamefont
  {Avouris}}, \bibinfo {author} {\bibfnamefont {S.-H.}\ \bibnamefont {Oh}},\
  and\ \bibinfo {author} {\bibfnamefont {T.}~\bibnamefont {Low}},\ }\bibfield
  {title} {\bibinfo {title} {Twisted two-dimensional material stacks for
  polarization optics},\ }\href@noop {} {\bibfield  {journal} {\bibinfo
  {journal} {Phys. Rev. Lett.}\ }\textbf {\bibinfo {volume} {128}},\ \bibinfo
  {pages} {193902} (\bibinfo {year} {2022})}\BibitemShut {NoStop}%
\bibitem [{\citenamefont {Zhang}\ \emph {et~al.}(2022)\citenamefont {Zhang},
  \citenamefont {Wang}, \citenamefont {Chen}, \citenamefont {Wen},\ and\
  \citenamefont {Luo}}]{zhang2022photonic}%
  \BibitemOpen
  \bibfield  {author} {\bibinfo {author} {\bibfnamefont {W.}~\bibnamefont
  {Zhang}}, \bibinfo {author} {\bibfnamefont {Y.}~\bibnamefont {Wang}},
  \bibinfo {author} {\bibfnamefont {S.}~\bibnamefont {Chen}}, \bibinfo {author}
  {\bibfnamefont {S.}~\bibnamefont {Wen}},\ and\ \bibinfo {author}
  {\bibfnamefont {H.}~\bibnamefont {Luo}},\ }\bibfield  {title} {\bibinfo
  {title} {Photonic spin hall effect in twisted few-layer anisotropic
  two-dimensional atomic crystals},\ }\href@noop {} {\bibfield  {journal}
  {\bibinfo  {journal} {Physical Review A}\ }\textbf {\bibinfo {volume}
  {105}},\ \bibinfo {pages} {043507} (\bibinfo {year} {2022})}\BibitemShut
  {NoStop}%
\bibitem [{\citenamefont {\v{S}i\v{s}kins}\ \emph {et~al.}(2019)\citenamefont
  {\v{S}i\v{s}kins}, \citenamefont {Lee}, \citenamefont {Alijani},
  \citenamefont {Van~Blankenstein}, \citenamefont {Davidovikj}, \citenamefont
  {Van Der~Zant},\ and\ \citenamefont {Steeneken}}]{Siskins2019}%
  \BibitemOpen
  \bibfield  {author} {\bibinfo {author} {\bibfnamefont {M.}~\bibnamefont
  {\v{S}i\v{s}kins}}, \bibinfo {author} {\bibfnamefont {M.}~\bibnamefont
  {Lee}}, \bibinfo {author} {\bibfnamefont {F.}~\bibnamefont {Alijani}},
  \bibinfo {author} {\bibfnamefont {M.~R.}\ \bibnamefont {Van~Blankenstein}},
  \bibinfo {author} {\bibfnamefont {D.}~\bibnamefont {Davidovikj}}, \bibinfo
  {author} {\bibfnamefont {H.~S.}\ \bibnamefont {Van Der~Zant}},\ and\ \bibinfo
  {author} {\bibfnamefont {P.~G.}\ \bibnamefont {Steeneken}},\ }\bibfield
  {title} {\bibinfo {title} {Highly anisotropic mechanical and optical
  properties of 2d layered as2s3 membranes},\ }\href@noop {} {\bibfield
  {journal} {\bibinfo  {journal} {ACS Nano}\ }\textbf {\bibinfo {volume}
  {13}},\ \bibinfo {pages} {10845} (\bibinfo {year} {2019})}\BibitemShut
  {NoStop}%
\bibitem [{\citenamefont {Shubnic}\ \emph {et~al.}(2020)\citenamefont
  {Shubnic}, \citenamefont {Polozkov}, \citenamefont {Shelykh},\ and\
  \citenamefont {Iorsh}}]{Shubnic2020}%
  \BibitemOpen
  \bibfield  {author} {\bibinfo {author} {\bibfnamefont {A.~A.}\ \bibnamefont
  {Shubnic}}, \bibinfo {author} {\bibfnamefont {R.~G.}\ \bibnamefont
  {Polozkov}}, \bibinfo {author} {\bibfnamefont {I.~A.}\ \bibnamefont
  {Shelykh}},\ and\ \bibinfo {author} {\bibfnamefont {I.~V.}\ \bibnamefont
  {Iorsh}},\ }\bibfield  {title} {\bibinfo {title} {High refractive index and
  extreme biaxial optical anisotropy of rhenium diselenide for applications in
  all-dielectric nanophotonics},\ }\href
  {https://doi.org/doi:10.1515/nanoph-2020-0416} {\bibfield  {journal}
  {\bibinfo  {journal} {Nanophotonics}\ }\textbf {\bibinfo {volume} {9}},\
  \bibinfo {pages} {4737} (\bibinfo {year} {2020})}\BibitemShut {NoStop}%
\bibitem [{\citenamefont {{\'A}lvarez-P{\'e}rez}\ \emph
  {et~al.}(2020)\citenamefont {{\'A}lvarez-P{\'e}rez}, \citenamefont {Folland},
  \citenamefont {Errea}, \citenamefont {Taboada-Guti{\'e}rrez}, \citenamefont
  {Duan}, \citenamefont {Mart{\'\i}n-S{\'a}nchez}, \citenamefont
  {Tresguerres-Mata}, \citenamefont {Matson}, \citenamefont {Bylinkin},
  \citenamefont {He} \emph {et~al.}}]{alvarez2020infrared}%
  \BibitemOpen
  \bibfield  {author} {\bibinfo {author} {\bibfnamefont {G.}~\bibnamefont
  {{\'A}lvarez-P{\'e}rez}}, \bibinfo {author} {\bibfnamefont {T.~G.}\
  \bibnamefont {Folland}}, \bibinfo {author} {\bibfnamefont {I.}~\bibnamefont
  {Errea}}, \bibinfo {author} {\bibfnamefont {J.}~\bibnamefont
  {Taboada-Guti{\'e}rrez}}, \bibinfo {author} {\bibfnamefont {J.}~\bibnamefont
  {Duan}}, \bibinfo {author} {\bibfnamefont {J.}~\bibnamefont
  {Mart{\'\i}n-S{\'a}nchez}}, \bibinfo {author} {\bibfnamefont {A.~I.}\
  \bibnamefont {Tresguerres-Mata}}, \bibinfo {author} {\bibfnamefont {J.~R.}\
  \bibnamefont {Matson}}, \bibinfo {author} {\bibfnamefont {A.}~\bibnamefont
  {Bylinkin}}, \bibinfo {author} {\bibfnamefont {M.}~\bibnamefont {He}}, \emph
  {et~al.},\ }\bibfield  {title} {\bibinfo {title} {Infrared permittivity of
  the biaxial van der waals semiconductor $\alpha$-moo3 from near-and far-field
  correlative studies},\ }\href@noop {} {\bibfield  {journal} {\bibinfo
  {journal} {Adv. Mater.}\ }\textbf {\bibinfo {volume} {32}},\ \bibinfo {pages}
  {1908176} (\bibinfo {year} {2020})}\BibitemShut {NoStop}%
\bibitem [{\citenamefont {Ma}\ \emph {et~al.}(2018)\citenamefont {Ma},
  \citenamefont {Alonso-Gonz{\'a}lez}, \citenamefont {Li}, \citenamefont
  {Nikitin}, \citenamefont {Yuan}, \citenamefont {Mart{\'\i}n-S{\'a}nchez},
  \citenamefont {Taboada-Guti{\'e}rrez}, \citenamefont {Amenabar},
  \citenamefont {Li}, \citenamefont {V{\'e}lez} \emph {et~al.}}]{ma2018}%
  \BibitemOpen
  \bibfield  {author} {\bibinfo {author} {\bibfnamefont {W.}~\bibnamefont
  {Ma}}, \bibinfo {author} {\bibfnamefont {P.}~\bibnamefont
  {Alonso-Gonz{\'a}lez}}, \bibinfo {author} {\bibfnamefont {S.}~\bibnamefont
  {Li}}, \bibinfo {author} {\bibfnamefont {A.~Y.}\ \bibnamefont {Nikitin}},
  \bibinfo {author} {\bibfnamefont {J.}~\bibnamefont {Yuan}}, \bibinfo {author}
  {\bibfnamefont {J.}~\bibnamefont {Mart{\'\i}n-S{\'a}nchez}}, \bibinfo
  {author} {\bibfnamefont {J.}~\bibnamefont {Taboada-Guti{\'e}rrez}}, \bibinfo
  {author} {\bibfnamefont {I.}~\bibnamefont {Amenabar}}, \bibinfo {author}
  {\bibfnamefont {P.}~\bibnamefont {Li}}, \bibinfo {author} {\bibfnamefont
  {S.}~\bibnamefont {V{\'e}lez}}, \emph {et~al.},\ }\bibfield  {title}
  {\bibinfo {title} {In-plane anisotropic and ultra-low-loss polaritons in a
  natural van der waals crystal},\ }\href@noop {} {\bibfield  {journal}
  {\bibinfo  {journal} {Nature}\ }\textbf {\bibinfo {volume} {562}},\ \bibinfo
  {pages} {557} (\bibinfo {year} {2018})}\BibitemShut {NoStop}%
\bibitem [{\citenamefont {Duan}\ \emph {et~al.}(2021)\citenamefont {Duan},
  \citenamefont {Alvarez-Perez}, \citenamefont {Tresguerres-Mata},
  \citenamefont {Taboada-Gutierrez}, \citenamefont {Voronin}, \citenamefont
  {Bylinkin}, \citenamefont {Chang}, \citenamefont {Xiao}, \citenamefont {Liu},
  \citenamefont {Edgar}, \citenamefont {Martin}, \citenamefont {Volkov},
  \citenamefont {Hillenbrand}, \citenamefont {Martin-Sanchez}, \citenamefont
  {Nikitin},\ and\ \citenamefont {Alonso-Gonzalez}}]{Duan2021}%
  \BibitemOpen
  \bibfield  {author} {\bibinfo {author} {\bibfnamefont {J.}~\bibnamefont
  {Duan}}, \bibinfo {author} {\bibfnamefont {G.}~\bibnamefont {Alvarez-Perez}},
  \bibinfo {author} {\bibfnamefont {A.~I.~F.}\ \bibnamefont
  {Tresguerres-Mata}}, \bibinfo {author} {\bibfnamefont {J.}~\bibnamefont
  {Taboada-Gutierrez}}, \bibinfo {author} {\bibfnamefont {K.~V.}\ \bibnamefont
  {Voronin}}, \bibinfo {author} {\bibfnamefont {A.}~\bibnamefont {Bylinkin}},
  \bibinfo {author} {\bibfnamefont {B.}~\bibnamefont {Chang}}, \bibinfo
  {author} {\bibfnamefont {S.}~\bibnamefont {Xiao}}, \bibinfo {author}
  {\bibfnamefont {S.}~\bibnamefont {Liu}}, \bibinfo {author} {\bibfnamefont
  {J.~H.}\ \bibnamefont {Edgar}}, \bibinfo {author} {\bibfnamefont {J.~I.}\
  \bibnamefont {Martin}}, \bibinfo {author} {\bibfnamefont {V.~S.}\
  \bibnamefont {Volkov}}, \bibinfo {author} {\bibfnamefont {R.}~\bibnamefont
  {Hillenbrand}}, \bibinfo {author} {\bibfnamefont {J.}~\bibnamefont
  {Martin-Sanchez}}, \bibinfo {author} {\bibfnamefont {A.~Y.}\ \bibnamefont
  {Nikitin}},\ and\ \bibinfo {author} {\bibfnamefont {P.}~\bibnamefont
  {Alonso-Gonzalez}},\ }\bibfield  {title} {\bibinfo {title} {Planar refraction
  and lensing of highly confined polaritons in anisotropic media},\ }\href
  {https://doi.org/10.1038/s41467-021-24599-3} {\bibfield  {journal} {\bibinfo
  {journal} {Nat. Commun.}\ }\textbf {\bibinfo {volume} {12}},\ \bibinfo
  {pages} {4325} (\bibinfo {year} {2021})}\BibitemShut {NoStop}%
\bibitem [{\citenamefont {Hu}\ \emph {et~al.}(2020)\citenamefont {Hu},
  \citenamefont {Ou}, \citenamefont {Si}, \citenamefont {Wu}, \citenamefont
  {Wu}, \citenamefont {Dai}, \citenamefont {Krasnok}, \citenamefont {Mazor},
  \citenamefont {Zhang}, \citenamefont {Bao}, \citenamefont {Qiu},\ and\
  \citenamefont {Alu}}]{Hu2020}%
  \BibitemOpen
  \bibfield  {author} {\bibinfo {author} {\bibfnamefont {G.}~\bibnamefont
  {Hu}}, \bibinfo {author} {\bibfnamefont {Q.}~\bibnamefont {Ou}}, \bibinfo
  {author} {\bibfnamefont {G.}~\bibnamefont {Si}}, \bibinfo {author}
  {\bibfnamefont {Y.}~\bibnamefont {Wu}}, \bibinfo {author} {\bibfnamefont
  {J.}~\bibnamefont {Wu}}, \bibinfo {author} {\bibfnamefont {Z.}~\bibnamefont
  {Dai}}, \bibinfo {author} {\bibfnamefont {A.}~\bibnamefont {Krasnok}},
  \bibinfo {author} {\bibfnamefont {Y.}~\bibnamefont {Mazor}}, \bibinfo
  {author} {\bibfnamefont {Q.}~\bibnamefont {Zhang}}, \bibinfo {author}
  {\bibfnamefont {Q.}~\bibnamefont {Bao}}, \bibinfo {author} {\bibfnamefont
  {C.-W.}\ \bibnamefont {Qiu}},\ and\ \bibinfo {author} {\bibfnamefont
  {A.}~\bibnamefont {Alu}},\ }\bibfield  {title} {\bibinfo {title} {Topological
  polaritons and photonic magic angles in twisted $\alpha$-moo$_3$ bilayers},\
  }\href {https://doi.org/10.1038/s41586-020-2359-9} {\bibfield  {journal}
  {\bibinfo  {journal} {Nature}\ }\textbf {\bibinfo {volume} {582}},\ \bibinfo
  {pages} {209–213} (\bibinfo {year} {2020})}\BibitemShut {NoStop}%
\bibitem [{\citenamefont {Duan}\ \emph {et~al.}(2020)\citenamefont {Duan},
  \citenamefont {Capote-Robayna}, \citenamefont {Taboada-Gutierrez},
  \citenamefont {Alvarez-Perez}, \citenamefont {Prieto}, \citenamefont
  {Martin-Sanchez}, \citenamefont {Nikitin},\ and\ \citenamefont
  {Alonso-Gonzalez}}]{Duan2020}%
  \BibitemOpen
  \bibfield  {author} {\bibinfo {author} {\bibfnamefont {J.}~\bibnamefont
  {Duan}}, \bibinfo {author} {\bibfnamefont {N.}~\bibnamefont
  {Capote-Robayna}}, \bibinfo {author} {\bibfnamefont {J.}~\bibnamefont
  {Taboada-Gutierrez}}, \bibinfo {author} {\bibfnamefont {G.}~\bibnamefont
  {Alvarez-Perez}}, \bibinfo {author} {\bibfnamefont {I.}~\bibnamefont
  {Prieto}}, \bibinfo {author} {\bibfnamefont {J.}~\bibnamefont
  {Martin-Sanchez}}, \bibinfo {author} {\bibfnamefont {A.}~\bibnamefont
  {Nikitin}},\ and\ \bibinfo {author} {\bibfnamefont {P.}~\bibnamefont
  {Alonso-Gonzalez}},\ }\bibfield  {title} {\bibinfo {title} {Twisted
  nano-optics: Manipulating light at the nanoscale with twisted phonon
  polaritonic slabs},\ }\href {https://doi.org/10.1021/acs.nanolett.0c01673}
  {\bibfield  {journal} {\bibinfo  {journal} {Nano Lett.}\ }\textbf {\bibinfo
  {volume} {20}},\ \bibinfo {pages} {5323–5329} (\bibinfo {year}
  {2020})}\BibitemShut {NoStop}%
\bibitem [{\citenamefont {Slavich}\ \emph {et~al.}(2023)\citenamefont
  {Slavich}, \citenamefont {Ermolaev}, \citenamefont {Tatmyshevskiy},
  \citenamefont {Toksumakov}, \citenamefont {Matveeva}, \citenamefont
  {Grudinin}, \citenamefont {Mazitov}, \citenamefont {Kravtsov}, \citenamefont
  {Syuy}, \citenamefont {Tsymbarenko} \emph {et~al.}}]{slavich2023exploring}%
  \BibitemOpen
  \bibfield  {author} {\bibinfo {author} {\bibfnamefont {A.~S.}\ \bibnamefont
  {Slavich}}, \bibinfo {author} {\bibfnamefont {G.~A.}\ \bibnamefont
  {Ermolaev}}, \bibinfo {author} {\bibfnamefont {M.~K.}\ \bibnamefont
  {Tatmyshevskiy}}, \bibinfo {author} {\bibfnamefont {A.~N.}\ \bibnamefont
  {Toksumakov}}, \bibinfo {author} {\bibfnamefont {O.~G.}\ \bibnamefont
  {Matveeva}}, \bibinfo {author} {\bibfnamefont {D.~V.}\ \bibnamefont
  {Grudinin}}, \bibinfo {author} {\bibfnamefont {A.}~\bibnamefont {Mazitov}},
  \bibinfo {author} {\bibfnamefont {K.~V.}\ \bibnamefont {Kravtsov}}, \bibinfo
  {author} {\bibfnamefont {A.~V.}\ \bibnamefont {Syuy}}, \bibinfo {author}
  {\bibfnamefont {D.~M.}\ \bibnamefont {Tsymbarenko}}, \emph {et~al.},\
  }\bibfield  {title} {\bibinfo {title} {Exploring van der waals materials with
  high anisotropy: geometrical and optical approaches},\ }\href@noop {}
  {\bibfield  {journal} {\bibinfo  {journal} {arXiv preprint arXiv:2309.01989}\
  } (\bibinfo {year} {2023})}\BibitemShut {NoStop}%
\bibitem [{\citenamefont {New}(2013)}]{new2013biaxial}%
  \BibitemOpen
  \bibfield  {author} {\bibinfo {author} {\bibfnamefont {G.}~\bibnamefont
  {New}},\ }\bibfield  {title} {\bibinfo {title} {Biaxial media revisited},\
  }\href@noop {} {\bibfield  {journal} {\bibinfo  {journal} {European Journal
  of Physics}\ }\textbf {\bibinfo {volume} {34}},\ \bibinfo {pages} {1263}
  (\bibinfo {year} {2013})}\BibitemShut {NoStop}%
\bibitem [{\citenamefont {Joannopoulos}(1995)}]{Molding1995photonic}%
  \BibitemOpen
  \bibfield  {author} {\bibinfo {author} {\bibfnamefont {J.}~\bibnamefont
  {Joannopoulos}},\ }\href@noop {} {\emph {\bibinfo {title} {Photonic crystals:
  Molding the flow of light}}}\ (\bibinfo  {publisher} {Pinceton Univ. Press},\
  \bibinfo {year} {1995})\BibitemShut {NoStop}%
\bibitem [{\citenamefont {Lipkin}(1964)}]{Lipkin1964}%
  \BibitemOpen
  \bibfield  {author} {\bibinfo {author} {\bibfnamefont {D.~M.}\ \bibnamefont
  {Lipkin}},\ }\bibfield  {title} {\bibinfo {title} {{Existence of a New
  Conservation Law in Electromagnetic Theory}},\ }\href
  {http://aip.scitation.org/doi/10.1063/1.1704165} {\bibfield  {journal}
  {\bibinfo  {journal} {J. Math. Phys.}\ }\textbf {\bibinfo {volume} {5}},\
  \bibinfo {pages} {696} (\bibinfo {year} {1964})}\BibitemShut {NoStop}%
\bibitem [{\citenamefont {Caloz}\ and\ \citenamefont
  {Sihvola}(2020)}]{caloz2020electromagnetic}%
  \BibitemOpen
  \bibfield  {author} {\bibinfo {author} {\bibfnamefont {C.}~\bibnamefont
  {Caloz}}\ and\ \bibinfo {author} {\bibfnamefont {A.}~\bibnamefont
  {Sihvola}},\ }\bibfield  {title} {\bibinfo {title} {Electromagnetic
  chirality, part 1: The microscopic perspective [electromagnetic
  perspectives]},\ }\href@noop {} {\bibfield  {journal} {\bibinfo  {journal}
  {IEEE Antennas and Propagation Magazine}\ }\textbf {\bibinfo {volume} {62}},\
  \bibinfo {pages} {58} (\bibinfo {year} {2020})}\BibitemShut {NoStop}%
\bibitem [{\citenamefont {Menzel}\ \emph {et~al.}(2010)\citenamefont {Menzel},
  \citenamefont {Rockstuhl},\ and\ \citenamefont {Lederer}}]{Menzel2010}%
  \BibitemOpen
  \bibfield  {author} {\bibinfo {author} {\bibfnamefont {C.}~\bibnamefont
  {Menzel}}, \bibinfo {author} {\bibfnamefont {C.}~\bibnamefont {Rockstuhl}},\
  and\ \bibinfo {author} {\bibfnamefont {F.}~\bibnamefont {Lederer}},\
  }\bibfield  {title} {\bibinfo {title} {{Advanced Jones calculus for the
  classification of periodic metamaterials}},\ }\href
  {https://doi.org/10.1103/PhysRevA.82.053811} {\bibfield  {journal} {\bibinfo
  {journal} {Phys. Rev. A}\ }\textbf {\bibinfo {volume} {82}},\ \bibinfo
  {pages} {053811} (\bibinfo {year} {2010})}\BibitemShut {NoStop}%
\bibitem [{\citenamefont {Semnani}\ \emph {et~al.}(2020)\citenamefont
  {Semnani}, \citenamefont {Flannery}, \citenamefont {{Al Maruf}},\ and\
  \citenamefont {Bajcsy}}]{Semnani2020}%
  \BibitemOpen
  \bibfield  {author} {\bibinfo {author} {\bibfnamefont {B.}~\bibnamefont
  {Semnani}}, \bibinfo {author} {\bibfnamefont {J.}~\bibnamefont {Flannery}},
  \bibinfo {author} {\bibfnamefont {R.}~\bibnamefont {{Al Maruf}}},\ and\
  \bibinfo {author} {\bibfnamefont {M.}~\bibnamefont {Bajcsy}},\ }\bibfield
  {title} {\bibinfo {title} {{Spin-preserving chiral photonic crystal
  mirror}},\ }\href {http://dx.doi.org/10.1038/s41377-020-0256-5} {\bibfield
  {journal} {\bibinfo  {journal} {Light Sci. Appl.}\ }\textbf {\bibinfo
  {volume} {9}},\ \bibinfo {pages} {23} (\bibinfo {year} {2020})}\BibitemShut
  {NoStop}%
\end{thebibliography}%

\end{document}